\documentclass[aos,preprint]{imsart}

\RequirePackage[OT1]{fontenc}
\RequirePackage[leqno,cmex10]{amsmath}
\RequirePackage{amsthm}
\RequirePackage[numbers]{natbib}
\RequirePackage[colorlinks,citecolor=blue,urlcolor=blue]{hyperref}

\usepackage{mathrsfs,dsfont}
\usepackage{latexsym, stmaryrd}
\usepackage{amssymb}
\usepackage[english]{babel}
\usepackage[latin1]{inputenc}
\usepackage{color}
\usepackage{amsfonts}
\usepackage{bbm}
\usepackage{enumerate}
\RequirePackage{xspace}
\usepackage[ruled, section]{algorithm}
\usepackage{algpseudocode}
\usepackage{subfigure}
\usepackage {graphicx}
\usepackage {pgf}
\usepackage{yhmath}
\usepackage[normalem]{ulem}

\usepackage{multido}
\usepackage{pstricks,pst-plot,pstricks-add,pst-math}
\usepackage{mathtools}

\usepackage{enumerate}

\numberwithin{equation}{section}
\numberwithin{figure}{section}
\numberwithin{table}{section}
\theoremstyle{plain}
\newtheorem{thm}{Theorem}[section]

\newtheorem{cor}[thm]{Corollary}

\newtheorem{defn}[thm]{Definition}

\theoremstyle{definition}

\newcommand{\R}{ \mathds R}
\newcommand{\Z}{ \mathds Z}
\newcommand{\N}{ \mathds N}

\theoremstyle{remark}
\newtheorem{rem}[thm]{Remark}
\newcommand{\ol}[1]{\left\langle #1\right\rangle}
\newcommand{\ul}[1]{\underline{#1}}
\newcommand{\bkt}[1]{\left\llbracket #1\right\rrbracket}
\newcommand{\ple}{\preccurlyeq }

\newcommand{\Gd}{\boldsymbol\delta}

\renewcommand{\le}{\leqslant}
\renewcommand{\ge}{\geqslant}

\newcommand{\eps}{\varepsilon}

\newcommand{\norm}[1]{\left\Vert#1\right\Vert}

\def\qed{\hfill $\blacksquare$}
\let\da=\downarrow
\let\ua=\uparrow
\let\ga=\alpha \let\gb=\beta \let\gc=\gamma \let\gd=\delta \let\gee=\epsilon
     \let\gl=\lambda

\let\gC=\Gamma \let\gD=\Delta

\newcommand{\cB}{\mathcal{B}}\newcommand{\cC}{\mathcal{C}}

\newcommand{\cN}{\mathcal{N}}

\newcommand{\V}[1]{\ensuremath{\boldsymbol{#1}}\xspace}
\newcommand{\vone}{\mathbf{1}}

\newcommand{\vA}{\mathbf{A}}\newcommand{\vB}{\mathbf{B}}

\newcommand{\vI}{\mathbf{I}}\newcommand{\vJ}{\mathbf{J}}

\newcommand{\vP}{\mathbf{P}}

\newcommand{\vU}{\mathbf{U}}\newcommand{\vW}{\mathbf{W}}
\newcommand{\vX}{\mathbf{X}}\newcommand{\vZ}{\mathbf{Z}}

\newcommand{\vz}{\mathbf{z}}
\newcommand{\dR}{\mathds{R}}

\DeclareMathOperator{\E}{\mathds{E}}
\DeclareMathOperator{\pr}{\mathds{P}}
\DeclareMathOperator{\argmin}{argmin}

\newcommand{\sM}{\mathscr{M}}

\def\beq{ \begin{equation} }
\def\eeq{ \end{equation} }
\def\beqx{ \begin{equation*} }
\def\eeqx{ \end{equation*} }
\def\beqa{\begin{eqnarray}}
\def\eeqa{\end{eqnarray}}
\def\beqax{\begin{eqnarray*}}
	\def\eeqax{\end{eqnarray*}}

\newcommand{\fD}{\mathfrak{D}}

\newcommand{\Pro}{ \mathds P}

\newcommand{\gre}{\epsilon}

\usepackage[T3,T1]{fontenc}
\DeclareSymbolFont{tipa}{T3}{cmr}{m}{n}
\DeclareMathAccent{\inv}{\mathalpha}{tipa}{16}

\begin{document}
	
	\begin{frontmatter}
		\title{General Community Detection with Optimal Recovery Conditions for Multi-relational Sparse Networks with Dependent Layers}
		\runtitle{Community Detection for Multi-relational Sparse Networks}
		%\thankstext{T1}{Footnote to the title with the ``thankstext'' command.}
		
		\begin{aug}
			\author{\fnms{Sharmodeep} \snm{Bhattacharyya}\thanksref{m2, t2}\ead[label=e1]{bhattash@science.oregonstate.edu}}
			\and
			\author{\fnms{Shirshendu} \snm{Chatterjee}\thanksref{m1,t3}\ead[label=e2]{shirshendu@ccny.cuny.edu}}
			%\and
			%\author{\fnms{Third} \snm{Author}\thanksref{t1,m2}
			%\ead[label=e3]{third@somewhere.com}
			%\ead[label=u1,url]{http://www.foo.com}}
			
			%\thankstext{t1}{Some comment}
			\thankstext{t2}{Supported in part by NSF grant DMS-1160319}
			\thankstext{t3}{Supported in part by PSC-CUNY and Simons Foundation}
			\runauthor{Bhattacharyya and Chatterjee}
			
			\affiliation{City University of New York\thanksmark{m1}}
			\affiliation{Oregon State University\thanksmark{m2}}

			\address{Department of Statistics\\
				239 Weniger Hall\\
				Corvallis, OR, 97331\\
				\printead{e1}}
			\address{Department of Mathematics\\
				North Academic Center 8/133\\
				160 Convent Ave\\
				New York, NY, 10031\\
				\printead{e2}}
			
			%\address{Address of the Third author\\
			%Usually a few lines long\\
			%Usually a few lines long\\
			%\printead{e3}\\
			%\printead{u1}}
		\end{aug}
		
		\begin{abstract} 
			%\indent One of the most common and crucial aspects of many network data sets is the dependence of network link structure on time.
			%% or other attributes. This has led the researchers to study dynamic, time-evolving networks. 
			%In this work, we extend the existing (static) nonparametric latent variable models in the context of  time-varying networks, and thereby propose a class of dynamic network models. For some special cases of these models (namely the dynamic stochastic block model and dynamic degree corrected block model), which assume that there is a common community structure for all networks, we consider the problem of identifying the common community structure. We consider two extensions of the (standard) spectral clustering method for the dynamic network models, and give theoretical guarantee that the spectral clustering methods produce consistent community detection in case of both dynamic stochastic block model and dynamic degree-corrected block model.  The methods are shown to work under sufficiently mild conditions on the number of time snapshots to detect associative, dissociative and mixed community structures, even if all the individual networks are very sparse and most of the individual networks are below community detectability threshold. We reinforce the validity of the theoretical results via simulations too.
			\indent Multilayer and multiplex networks are becoming common network data sets in recent times. We consider the problem of identifying the common community structure for a special type of multilayer networks called multi-relational networks. We consider extensions of the spectral clustering methods for multi-relational networks and give theoretical guarantees that the spectral clustering methods recover community structure consistently for multi-relational networks generated from multilayer versions of both stochastic and degree-corrected block models even with dependence between network layers. The methods are shown to work under optimal conditions on the degree parameter of the networks to detect both assortative and disassortative community structures with vanishing error proportions even if individual layers of the multi-relational network has the network structures below community detectability threshold. We reinforce the validity of the theoretical results via simulations too.
			
		\end{abstract}
		
		
		
	\end{frontmatter}
	
	\section{Introduction}
	\label{sec_intro}
	
	Statistical analysis of network data has now become a well-studied field within statistics (see \citep{goldenberg2010survey, kolaczyk2014statistical} for reviews). Methods for network data analysis are being developed not only in the discipline of statistics but also in computer science, physics, and mathematics. Network datasets show up in several disciplines. Examples include networks originating from biosciences such as gene regulation networks \cite{emmert2014gene}, protein-protein interaction networks \cite{de2010protein}, structural \cite{rubinov2010complex} and functional networks \cite{friston2011functional} of brain and epidemiological networks \cite{reis2007epidemiological}; networks originating from social media such as Facebook, Twitter and LinkedIn \cite{faloutsos2010online}; citation and collaboration networks \cite{lehmann2003citation}; information and technological networks such as internet-based networks \cite{adamic2005political}, power networks \cite{pagani2013power} and cell-tower networks \cite{isaacman2011identifying}.
	%Most of the research in the statistics community focuses on developing methods for addressing statistical inference questions based on a single observed network as data. We will refer to such single networks as {\it static networks} in this paper. 
	There are several active areas of research in developing statistical inference methods for network data analysis and also deriving the theoretical properties of the statistical methods. Examples of inferential questions that have received a lot of attention in current research include fitting of random graph models to the network data sets \cite{goldenberg2010survey}, finding stochastic properties of summary statistics of networks like subgraph counts \cite{bickel2011method}, community detection \cite{fortunato2010community} and link prediction \cite{liben2007link}. 
	
	In this paper, we focus on the problem of recovering a common community structure present in a finite sequence of (possibly sparse) networks. The community detection problem can be thought of as a  \emph{vertex clustering problem}, in which the goal is to divide the set of vertices of a given network (or a finite sequence of networks)  into groups based on some common properties of the vertices. The main goal in community detection is to partition the vertices of a graph (or a finite sequence of graphs)  into groups such that the average numbers of connections within the groups are \emph{significantly different} than that between groups. Communities in networks are usually called \emph{assortative} (see \textsection \ref{sec_model} for more details) if the average number of connections within communities is \emph{significantly higher} than that between communities. Communities in networks are usually called \emph{disassortative}, if the average number of connections within communities is \emph{significantly lesser} than the average number of connections between communities. A network may consist of both assortative and disassortative communities (see \cite{newman2003mixing, newman2004finding}). Since many works on community detection only deal with assortative community detection, to avoid ambiguity we have referred our goal as \emph{general} community detection. \emph{In this paper, we do not restrict ourselves to any specific type of community structure}.
	%Communities in networks are usually called \emph{dissociative}, if the average number of connections within communities is \emph{significantly lesser} than the average number of connections between communities. A network may consist of both associative and dissociative communities. 
	%In this paper, we focus on finding the common community structure, when the communities are assortative.  We will treat the case when communities are not necessarily assortative in a future paper.
	%Our research results for finding communities which are not assortative will appear in a different venue.
	
	Several random graph models have been proposed in the literature with a mathematically rigorous definition of community labels for vertices. Examples of random graph models for a single network with community structure include stochastic block models \cite{holland1983stochastic}, degree-corrected block models \cite{karrer2011stochastic} and random dot product models \cite{young2007random}.
	%, exchangeable network models \cite{bickel2009nonparametric}). 
	Many methods have been proposed in the statistics and machine learning literature to recover community labels (see \cite{fortunato2010community} for a review) for a given single network. The methods can be broadly classified into two types, namely (i) \emph{model-based approaches} (e.g.,~different likelihood-based methods \cite{bickel2009nonparametric}), where the methods are developed assuming a specific generative model for the given network, and (ii) model agnostic approaches (e.g.~modularity based methods \cite{newman2004finding}, spectral clustering methods \cite{MR2893856}, label propagation \cite{gregory2010finding}), where the methods are developed without a specific generative model in mind.  
	
	Most of the research on network data in statistics literature has focused on questions based on a single observed network as data. However, multiple network datasets (a finite sequence of networks) are currently becoming common in many applications. Examples of applications include, neuroscience \cite{bassett2017network, thompson2017static}, economics \cite{bargigli2015multiplex}, sociology and social networks \cite{heaney2014multiplex, lewis2012social}, ecology \cite{pilosof2017multilayer}, epidemiology \cite{zhao2014immunization}, and technological networks \cite{sen2014identification, zignani2014exploiting}. Depending on the structure and interconnectivity among a finite sequence of networks, various kinds of multiple networks have been considered in the literature, e.g.,~multilayer networks, multiplex networks, multi-relational networks, multidimensional networks, time-evolving networks, dynamic networks, and hypergraphs \cite{boccaletti2014structure, kivela2014multilayer}. A \emph{multi-relational network} consists of a finite set of networks (each such network is called a \emph{network layer}) having the same vertex set but possibly different edge sets in different layers. Temporal networks having the same vertex set and time-evolving edge sets can also be considered as \emph{multi-relational networks}. \emph{We consider the problem of community detection based on \emph{multi-relational} network datasets, which is a generalization of its analog for a single-layer network}.  
	
	Community detection using the spectral decomposition of matrices associated with graphs is a common statistical method. Spectral clustering has several advantages - firstly, the method is model agnostic. Secondly, spectral clustering is highly scalable, as scalable implementations of matrix factorization algorithms is an active research topic in the numerical analysis literature \cite{blackford1997scalapack}. Thirdly, spectral clustering methods have also been shown to work in recovering community labels for single-layer networks under various probabilistic models and analyzed in many subsequent papers (see \cite{shi2000normalized}, \cite{ng2002spectral}, \cite{MR2396807}, \cite{MR2893856}, \cite{sussman2012consistent}, \cite{lei2015consistency}, \cite{bhattacharyya2014community}, \cite{gao2017achieving}). Also for a single-layer network, many of the proposed community detection methods \cite{chin2015stochastic, joseph2016impact, abbe2017community, gao2017achieving, le2017concentration, gao2018community} in the literature has been shown to recover community labels for sparse networks, but still the scalability of the methods have rarely been addressed. 
	
	Most of the statistical and probabilistic models for multiple networks that appear in the literature are extensions of random graph models for a single network into the multiple networks setup.  Examples of such models include extension of latent space models \cite{sarkar2005dynamic}, \cite{sewell2014latent}, mixed membership block models \cite{ho2011evolving}, random dot-product models \cite{tang2013attribute}, stochastic block models \cite{xu2014dynamic}, \cite{xu2015stochastic}, \cite{matias2017statistical}, \cite{ghasemian2016detectability}, \cite{corneli2016exact}, \cite{zhang2017random}, \cite{pensky2019dynamic}, and Erd\'{o}s-R\'{e}nyi graph models \cite{crane2015time}. Also, some Bayesian models and associated inference procedures have been proposed in the context of multiple networks  \cite{yang2011detecting}, \cite{durante2014nonparametric}. \emph{In this paper, for theoretical analysis we have considered a multilayer version of stochastic and degree-corrected block models which has been used in some of the previous works \cite{han2015consistent}.}
	
	Several recent works have focused on developing statistical inference procedures based on different versions of multilayer networks \cite{xu2014dynamic, han2015consistent, matias2017statistical, zhang2017random, paul2016consistent}. Some model-agnostic methods have also been proposed to detect communities in multilayer networks \cite{tang2009clustering, kumar2010co, dong2012clustering, chen2017multilayer, paul2017spectral, pensky2019spectral}. However, only a few of the recently proposed algorithms \cite{han2015consistent, paul2016consistent, taylor2016enhanced, paul2017spectral, chen2017multilayer, pensky2019spectral} attempts to evaluate the performance of the proposed community recovery procedures theoretically when the multilayer network is sampled from some random network generating model. None of the proposed methods have been proven to work for multilayer networks in which an aggregation of individual networks is sparse, namely when the total degree of a typical vertex in the aggregated network goes to infinity arbitrarily slowly. Also, some recent works like \cite{mercado2018power}, considers power of Laplacian matrices for community detection in multilayer networks but they only consider networks generated from special cases of multilayer stochastic block model. \emph{So, to the best of our knowledge, no known polynomial-time community detection algorithm with proven theoretical guarantee to consistently recover community labels within a general class of sparse multilayer networks has been proposed.} \emph{Also, to the best of our knowledge, none of the recently proposed community detection algorithms in the literature have been shown to recover community labels under general dependence structures between the network layers. }
	
	\subsection{Contributions of our work} 
	\label{sec_contribution}
	We address some of the limitations of current works in this paper, so, we propose and analyze two spectral clustering algorithms for finding the common community structure within a given finite sequence of networks with possible dependence structures. The proposed algorithms are naturally scalable and model agnostic, and they work for a single network as well as for multilayer networks, irrespective of edge density of individual networks as well as their aggregated versions. To evaluate the performance of the proposed community recovery algorithms theoretically and see when they perform consistently, we consider a particular case of multilayer networks, \emph{multi-relational networks} \cite{cai2005community, kivela2014multilayer} generated from a multilayer generalization of stochastic and degree-corrected block models \cite{han2015consistent}.  
	
	The main contributions of our work are the following.
	\begin{itemize}
		\item[(a)] We propose two novel methods based on spectral clustering of sum of squared adjacency matrices for recovering community labels in multi-relational networks with a common community structure. The methods can be used for community detection in a single-layer network too. 
		\item[(b)] We also prove analytically that, under the mildest (necessary) parametric conditions, the proposed spectral clustering methods identify communities in the networks generated from single-layer or multilayer stochastic block models and degree-corrected block models consistently. We show analytically that in the multi-relational networks generated from multilayer versions of stochastic and degree-corrected block models, our spectral clustering methods can recover the common community structure consistently even if each of the individual network layers has fixed size and is highly sparse (e.g., has a constant average degree) and has connectivity below the community detectability threshold as long as the aggregated version of the network satisfies certain conditions. 
		\item[(c)] It has been theoretically shown that the proposed community detection methods are flexible enough to work for both sparse and dense networks. It has been theoretically shown that the methods are flexible enough to identify both assortative and disassortative community structures even when the community structures vary between layers.
		\item[(d)] It has been theoretically shown that the proposed community detection methods recover community labels even in the presence of dependence between network layers.
		\item[(e)] We also propose a method for detecting the number of communities in the multi-relational networks. The proposed method has been shown to recover the correct number of communities asymptotically. 		
	\end{itemize}
	
	\subsection{Structure of the paper} 
	\label{sec_organization}
	The remainder of the paper is organized as follows. In \textsection \ref{sec_model}, we introduce the multiple network models. In \textsection \ref{sec_method}, we describe the spectral clustering methods.  In \textsection \ref{sec_theory}, we state the theoretical results regarding the performance of the proposed spectral clustering methods. 
%	We also give a sketch of the proofs but the detailed proofs are given in the Appendix \cite{bhattacharyya2020supplement}. 
	In \textsection \ref{sec_simulation}, we demonstrate the effectiveness of the methods for simulated datasets. 
	
	\section{Multi-relational Network Data and Model}
	\label{sec_model}
	
	\subsection{Multi-relational networks data}
	\label{sec_multiple}
	In this paper, we suppose that the observed data consists of a single network or a \emph{multi-relational network}. The formal definition of a \emph{multi-relational network} is given below. 
	\begin{defn}[Multi-relational network]
		A \emph{multi-relational network} consists of a finite sequence of unlabeled graphs $\{G_n^{(t)}; t=1, \ldots, T\}$ on the same vertex set $V_n=\{v_1, v_2, \ldots, v_n\}$ having $n$ vertices but the edge sets of the graphs  may be different. $G_n^{(t)}$ is referred as the the $t$-th \emph{network layer}. 
		%So, the vertex sets of the graphs are the same,
		%	Note that the vertex sets $V(G_n^{(t)})$ of $G_n^{(t)}$ don't change with $t$, and 
		%. We will refer to $G_n^{(t)}$ as the \emph{network layer} at instance $t$.
	\end{defn}
	A \emph{multi-relational network} can also be considered as an \emph{edge-colored multi-graph}, where different colors correspond to edge sets of different network layers.
	The $t$-th layer $G_n^{(t)}$ is represented by the corresponding adjacency matrix $\vA^{(t)}_{n\times n}$ whose elements are $\vA^{(t)}_{ij}\in\{0,1\}$. $\vA^{(t)}_{ij} = 1$ if node $v_i$ is linked to node $v_j$ at time $t$, and $\vA^{(t)}_{ij} = 0$ otherwise. Thus, the numerical data for the community detection problem consists of $T\ge 1$ adjacency matrices $\left\{\vA^{(1)}_{n\times n}, \ldots, \vA^{(T)}_{n\times n}\right\}$.
	%We will refer to the network model of a specific network snapshot as the \emph{single network model}.
	We shall only consider undirected and unweighted graphs in this paper. However, the conclusions of the paper can be extended to positively weighted graphs with non-random weights in a quite straightforward way by considering weighted adjacency matrices. The theoretical analysis in this paper can easily be extended to positively weighted adjacency matrices. Also, in this paper we consider that the multi-relational network has a \emph{common community structure}. So, the multi-relational network $\{G_n^{(t)}\}_{t=1}^T$ has the \emph{same} community structure in every layer with $K$ as the number of communities. Let us denote $\vZ_{n\times K}$ to be the actual common community membership matrix of the nodes in each of the graphs $G_n^{(t)}$, where, $\vZ_{ik} = 1$ if the $i$-th node belongs to the $k$-th community for all $G_n^{(t)}$ and zero otherwise. 
	
	\subsection{Notations}
	\label{sec_notation}
	Let $[n] := \{1, 2, \ldots, n\}$ for any positive integer $n$, $\sM_{m,n}$ be the set of all $m\times n$ matrices which have exactly one 1 and $n-1$ 0's in each row. $\R^{m\times n}$ denotes the set of all $m\times n$ real matrices. $||\cdot||_2$ is used to denote Euclidean $\ell_2$-norm for vectors in $\R^{m\times 1}$. $||\cdot||$ is the spectral norm on $\R^{m\times n}$. $||\cdot||_F$ is the Frobenius norm on $\R^{m\times n}$, namely $||M||_F := \sqrt{trace(M^T M)}$. $\vone_{n} \in \R^{n\times 1}$ consists of all 1's, $\mathbf 1_A$ denotes the indicator function of the event $A$. $\vI_n$ is the $n\times n$ identity matrix and $\vJ_n := \vone_n\vone_n^T$ is the $n\times n$ matrix of all 1's. For $\vA\in\dR^{n\times n}$, $\cC(\vA)$ and $\cN(\vA)$ denote its column space and null space of $\vA$ respectively, and $\gl_1(\vA), \gl_1^+(\vA)$ denote the smallest and smallest positive eigenvalues of $\vA$. 
	If $\vA\in\R^{m\times n}$,  $I\subset [m]$ and $j\in [n]$, then $\vA_{I,j}$ (resp.~$\vA_{I,*}$)  denotes the submatrix  of $\vA$ corresponding to row index set $I$ and column index $j$ (resp.~index set $[n]$). For $\vA\in \R^{n\times n}$,  $\ol{\vA}$ denotes the matrix $\vA$ with its diagonal zeroed out: $\ol{A}_{i,j}=A_{i,j}$ if $i \ne j, i,j\in [n]$ and $\ol{A}_{i,i} = 0$ for $i\in [n]$.
	%\beq \label{oldef}
	%\ol{A}_{i,j}=\begin{cases} A_{i,j} & \text{ if } i \ne j \\ 0 & \text{if } i=j \end{cases}.\eeq
	
	For a random variable (real valued or matrix valued)  $X$, we write $\bkt{X}:=X-\E(X)$.  For two random variables $X$ and $Y$, we write $X\ple Y$ to denote that $X$ is stochastically dominated by $Y$.   $\gl_i(\vW), i\in[n],$ will denote the $i$-th largest eigenvalue of $\vW\in\R^{n\times n}$. 
	
	\subsection{Multilayer Stochastic Block Model}
	The first model that we consider is an extension of stochastic block model (SBM) for generating multi-relational networks. We will refer to this model as {\it multilayer stochastic block model} (MSBM) in the paper. MSBM for $K$ communities ($\cC_1, \ldots, \cC_K$) can be described in terms of three parameters: (i) the membership vector $\V{z}=(z_1, \ldots, z_n)$, where each $z_i \in \{1, \ldots, K\}$; (ii)  the $K\times K$ connectivity probability matrices $\vB:=\left(\vB^{(t)}: 1\le t\le T\right)$ and (iii) the $K\times 1$ probability vector of allocation in each community, $\V{\pi} = (\pi_1, \ldots, \pi_K)$. The MSBM having parameters $(\V{z}, \V{\pi}, \V{B})$ is given by
	\begin{eqnarray}
	\label{eq_sbm0}
	\vz_1, \ldots, \vz_n & \stackrel{iid}{\sim} & \mbox{Mult}(1;(\pi_1,\ldots,\pi_K)),\\
	\label{eq_sbm1}
	\Pro\left(A^{(t)}_{ij} = 1 | \vz_i, \vz_j\right) & = & B^{(t)}_{\vz_i \vz_j}\ \ \ \text{for } i >j,\ i,j\in [n] . 
	%\label{eq_sbm2}
	%\Pro\left(A^{(t)}_{ij} = 1|\vz^t_i , \vz^t_j,\V{\gt} \right) & = & \gt_i\gt_jB^t_{\vz^t_i, \vz^t_j} 
	\end{eqnarray}
	
	Suppose $\vZ \in \sM_{n,K}$ denotes the actual membership matrix. $\vZ$ is unknown and we wish to estimate it. If for $i \in [n]$ the corresponding community index is $\vz_i \in [K]$, then clearly  
	\[ \vZ_{ij} = \mathbf 1_{\{\vz_i=j\}},\]
	In a MSBM$(\vz, \V{\pi}, \vB)$, independent edge formation is assumed given the edge probability matrices $\vP^{(t)}:=(P^{(t)}_{ij})_{i,j\in[n]}$. 
	%The link probability at time $t$ between two members from the $k$-th and $l$-th community respectively is given by $B^{(t)}_{kl}$, then 
	So, for $i, j \in [n]$ with $i \ne j$ and for $t \in [T]$
	\beq \label{A^t bmdef}
	A^{(t)}_{i,j} \sim Bernoulli(P^{(t)}_{i,j}), \text{ where }
	\vP^{(t)} := \vZ \vB^{(t)}\vZ^T.\eeq
	
	\subsection{Multilayer Degree Corrected Block Model}
	Multilayer degree-corrected block model is an extension of the degree corrected block model (DCBM) for generating multi-relational networks. The multilayer degree-corrected block model (MDCBM) for $K$ communities ($\cC_1, \ldots, \cC_K$) can be described in terms of four sets of parameters: (i) the membership vector $\V{z}=(z_1, \ldots, z_n)$, where each $z_i \in \{1, \ldots, K\}$,  (ii)  the $K\times K$ connectivity probability matrices $\vB:=\left(\vB^{(t)}: 1\le t\le T\right)$, (iii)
	a given set of \emph{degree parameters} $\V{\psi} = (\psi_1, \ldots, \psi_n)$ and (iv) the $K\times 1$ probability vector of allocation in each community, $\V{\pi}=(\pi_1, \ldots, \pi_K)$. The MDCBM having parameters $(\V{z}, \V{\pi}, \V{B}, \V{\psi})$ is given by
	\begin{eqnarray}
	\label{eq_dcbm0}
	\vz_1, \ldots, \vz_n & \stackrel{iid}{\sim} & \mbox{Mult}(1;(\pi_1,\ldots,\pi_K)),\\
	\label{eq_dcbm1}
	\Pro\left(A^{(t)}_{ij} = 1\right| \vz_i, \vz_j) & = & \psi_i\psi_jB^{(t)}_{\vz_i \vz_j} \ \ \ \text{for } i >j,\ i,j\in [n]. 
	%\label{eq_sbm2}
	%\Pro\left(A^{(t)}_{ij} = 1|\vz^t_i , \vz^t_j,\V{\gt} \right) & = & \gt_i\gt_jB^t_{\vz^t_i, \vz^t_j} 
	\end{eqnarray}
	
	The inclusion of $\V{\psi}$ involves the obvious issue of identifiability. In order to avoid this issue we assume that \cite{lei2015consistency} 
	\begin{align}
	\label{eq_dcbm_id}  
	\max_{i\in\cC_k} \psi_i=1 \text{ for all } k\in\{1, 2, \ldots, K\}.   
	\end{align}
		
	In an MDCBM$(\vz, \V{\psi}, \V{\pi}, \vB)$ also independent edge formation is assumed given the edge probability matrices $\tilde\vP^{(t)}$. Here also, for $i, j \in [n]$ with $i \ne j$ and for $t \in [T]$
	\beq \label{A^t def}
	A^{(t)}_{i,j} \sim Bernoulli(\tilde P^{(t)}_{i,j}), \text{ where }
	\tilde\vP^{(t)}  := \fD(\V{\psi})\vZ \vB^{(t)}\vZ^T\fD(\V{\psi})
	\eeq
	where, $\fD(\V{\psi}) = \text{diag}(\V{\psi})$.
	
	\subsection{Community Structure}
	\label{sec_comm}
	The \emph{assortative} and \emph{disassortative} community structures can be defined formally using the parameter structures of multilayer stochastic block models and degree-corrected block models, specially, the connectivity probability matrices $\{\vB^{(t)}\}_{t=1}^T$.
	\begin{defn} \label{AssoDisso}
		For a multi-relational network generated from MSBM or MDCBM with connectivity probability matrices $\{\vB^{(t)}\}_{t=1}^T$, the $t$-th layer is said to have -
		%\begin{itemize}
		(i) \emph{assortative} structure if all the eigenvalues of $\vB^{(t)}$ are positive;
		(ii) \emph{disassortative} structure if at least one of the eigenvalues of $\vB^{(t)}$ is negative.
		%\end{itemize}
	\end{defn}
	In this paper, we consider the case where, the community membership does not change between the layers of multi-relational network but the connectivity structure can change arbitrarily between layers and the layers can have either assortative or disassortative community structures.
	
	\section{Community Detection Algorithms}
	\label{sec_method}
	
	\subsection{Spectral clustering  using sum of squared adjacency matrices}
	\label{sec_algo_1}
	Let $\vZ \in \sM_{n,K}$ denote the actual community membership matrix of the nodes, where, if $\vZ_{ik} = 1$ ($i\in[n]$ and $k\in[K]$), then, node $i$ belongs to $k$-th community. The goal of the statistical methods is estimation of $\vZ$ based on the adjacency matrix data $\vA^{(1)}, \ldots, \vA^{(T)}$. We apply the spectral clustering method to a matrix which is derived from the \textbf{sum of the squared adjacency matrices} 
	$\vA_0^{[2]}:=\sum_{t\in[T]} \left(\vA^{(t)}\right)^2$. We zero out the diagonal of $\vA_0^{[2]}$ to obtain $\ol{\vA_0^{[2]}}$. 
	
	The squared adjacency matrices capture both assortative and disassortative community structures in different network layers. The squared adjacency matrices maintain the community structure in form of an assortative structure, since the non-zero elements of squared adjacency matrices represent paths of length two between the corresponding nodes. So, summing up squared adjacency matrices maintain both assortative and disassortative community structures in different network layers in an assortative form.
	
	Now, we prune $\langle\vA^{[2]}_0\rangle$ so that the empirical spectrum of the pruned matrix captures the community structure even if the networks are sparse. 
	For each node $i$, get the number of max-one-neighbors (resp.~total-two-neighbors) $D^{[1]}_i$ (resp.~$D^{[2]}_i$).  Also, get the average number of two-neighbors $\bar{d^2}$ of the nodes.
	\[
	D^{[1]}_i :=\max_{t\in[T]}\sum_{j\in[n]} A^{(t)}_{i,j}, D^{[2]}_i:=\sum_{ j\in[n]}\ol{A^{[2]}_0}_{i,j} \text{ for } i\in[n], \bar{d^2} = \frac{1}{nT}\sum_{i\in[n]}  D^{[2]}_i \]
	Then we sort the numbers $(D^{[l]}_i, i\in[n])$ to get the order statistics $D^{[l]}_{(1)}\le\cdots\le D^{[l]}_{(n)}$ for both $l=1, 2$. 
	Let $n'$ be the number of nodes and $1\le k_1<k_2<\cdots<k_{n'}\le n$ be the node indices having at most $D^{[1]}_{(n+1-\gC_1)}$ many max-one-neighbor and at most $D^{[2]}_{(n+1-\gC_2)}$  many total-two-neighbors, where
	\begin{align}
	\gC_1:=\left\lceil n\exp\left(-\frac 12T^{1/2}\left[\bar{d^2}\right]^{3/4}\right)\right\rceil, \gC_2:=\left\lceil n\exp\left(-\frac 13T\left[\bar{d^2}\right]^{1/2}\right)\right\rceil.  \label{gC def}
	\end{align}
	
	\vspace{0.2in}
	\framebox[\textwidth]
	{\centering\parbox{.95\textwidth}{
			\textbf{Algorithm 1:} Spectral Clustering of the Sum of  the Squared Adjacency Matrices \\
			\textbf{Input:} Adjacency matrices $\vA^{(1)}, \vA^{(2)}, \ldots, \vA^{(T)}$; number of communities $K$; approximation parameter $\gre$. \\ 
			%truncation parameter $\gd$.\\
			\textbf{Output:} Membership matrix $\hat\vZ_0$. \\
			\begin{enumerate}
				\item Obtain $\vA_0^{[2]} := \sum_{t=1}^T \left(\vA^{(t)}\right)^2$ (sum of squares of the adjacency matrices) and zero out its diagonal to get
				$\ol{\vA_0^{[2]}}$.
				\item Get~$D^{[1]}_i :=\max_{t=1}^T\sum_{j=1}^nA^{(t)}_{i,j}$~and~$D^{[2]}_i:=\sum_{ j=1}^n\langle A^{[2]}_0\rangle_{i,j}$~for~$ i\in[n]$. 
				\item Get the order statistics $D^{[1]}_{(1)} \le \cdots \le  D^{[1]}_{(n)}$ and $D^{[2]}_{(1)} \le \cdots \le D^{[2]}_{(n)}$.  
				\item Get $\bar{d^2} :=\frac{1}{nT} \sum_{i=1}^nD^{[2]}_i$. Get $\gC_1$ and $\gC_2$ as in \eqref{gC def}.
				%\[\gC_1:=\left\lceil n\exp\left(-\frac 12T^{1/2}\left[\bar{d^2}\right]^{3/4}\right)\right\rceil, \gC_2:=\left\lceil n\exp\left(-\frac 13T\left[\bar{d^2}\right]^{1/2}\right)\right\rceil. \]
				\item Get $\{i\in[n]: D^{[l]}_i \le D^{[l]}_{(n+1-\gC_l)} \text{ for both } l=1, 2\}$ and sort its entries in ascending order to have $1\le k_1<\cdots <k_{n'}\le n$.
				\item Get submatrix $\vA^{[2]}\in\R^{n'\times n'}$ of $\langle\vA_0^{[2]}\rangle$, where  $A^{[2]}_{i,j}=\langle\vA_0^{[2]}\rangle_{k_i,k_j}$. 
				\item Obtain $\hat\vU\in\R^{n'\times K}$ consisting of $K$ orthogonal eigenvectors of $\vA^{[2]}$ corresponding to its largest eigenvalues. 
				\item Use $(1+\gre)$ approximate $K$-means clustering algorithm on the row vectors of $\hat\vU$ to obtain $\hat\vZ \in \sM_{n',K}$ and $\hat\vX \in \R^{K \times K}$ satisfying \eqref{eq:kmeans}.
				\item Extend $\hat\vZ$ to obtain $\hat\vZ_0\in\sM_{n,K}$ as follows. $(\hat\vZ_0)_{i,*} = \hat\vZ_{j,*}$ (resp.~$(1,0,\ldots, 0)$) for $i=k_j$ (resp.~$i\notin\{k_1, \ldots, k_{n'}\}$). 
				\item $\hat\vZ_0$ is the estimate of $\vZ$.
			\end{enumerate}
	}}\\
	Let $\vA^{[2]}\in\R^{n'\times n'}$ be the submatrix of $\langle\vA_0^{[2]}\rangle$ such that $A^{[2]}_{i,j} := \langle A_0^{[2]}\rangle_{k_i,k_j}$ for $i, j\in [n']$. 
	Next, we obtain the leading $K$ eigenvectors of $\vA^{[2]}$ corresponding to its largest eigenvalues. Suppose $\hat\vU\in \R^{n'\times K}$ contains those eigenvectors as columns. Then, we use an $(1+\gre)$-approximate $K$-means clustering algorithm on the row vectors of $\hat\vU$ to obtain $\hat\vZ \in \sM_{n',K}$ and $\hat \vX \in \R^{K \times K}$ such that
	\begin{align} \label{eq:kmeans}
	\norm{\hat\vZ\hat\vX - \hat\vU}_F^2 \le (1+\gre) \min_{\V{\gC} \in \sM_{n'\times K}, \vX \in \R^{K\times K}}  \norm{\V{\gC} \vX - \hat\vU}_F^2.
	\end{align}
	Finally, $\hat\vZ$ is extended to $\hat\vZ_0\in\sM_{n,K}$ by taking $(\hat\vZ_0)_{k_j,*}:=\hat\vZ_{j,*}$ for all $j\in[n']$,  and filling in the remaining rows arbitrarily. One simple choice would be assigning all the pruned nodes to the first community.
	\[ (\hat\vZ_0)_{i,*} := \begin{cases} \hat\vZ_{j,*} & \text{ if  $i=k_j$ for some $j\in[n']$} \\ (e^K_1)^T & \text{otherwise}\end{cases}\]
	$\hat\vZ_0$ is the estimate of $\vZ$ from this method. 
	
	The reason for using an $(1+\gre)$-approximate $K$-means clustering algorithm is completely theoretical. $K$-means clustering is originally an NP-hard problem with any $K$-means clustering algorithm generating an approximate solution. However, we need a guarantee on the error of $K$-means clustering algorithm. So, we choose to use the $K$-means algorithms that can give us a guarantee on the error of the optimized objective function like algorithms proposed in \cite{kumar2004simple, feldman2007ptas}. 
	
	\subsection{Spherical Spectral Clustering Algorithm for Sum of Squared Adjacency Matrices} 
	\label{sec_algo_2}
	The goal is to estimate the community membership matrix $\vZ$ based on the adjacency matrices $\vA^{(1)}, \ldots, \vA^{(T)}$. We apply the spherical spectral clustering method, which is a modification of Algorithm 1. The modification is motivated from the works \cite{jin2015fast} and \cite{lei2015consistency}.
	Let $\vA_0^{[2]}$, $\vA^{[2]}$ and $\hat\vU$ be as in \textsection \ref{sec_algo_1}. 
	%Recall that $\hat\vU \in\R^{n'\times K}$ contains the leading $K$ eigenvectors (corresponding to the largest absolute eigenvalues) of $\vA$ as columns.  
	For $\hat\vU$, let $n''$ be the number of nonzero rows (with indices $1\le l_1 < l_2 <\cdots <l_{n''}\le n'$). Let $\hat\vU^+ \in \R^{n''\times K}$ consist of the normalized nonzero rows of $\hat\vU$, i.e.~$\hat\vU^+_{i,*}=(||\hat\vU_{l_i,*}||^{-1}_2)\hat\vU_{l_i,*}$ for $i\in [n'']$.  
	Apply an $(1+\gee)$ approximate $K$-means clustering algorithm on the rows of $\hat\vU^+$ to get $\check\vZ^+ \in \sM_{n'',K}$ 
	and $\check \vX \in \R^{K \times K}$~so~that
	\beq \label{k mean_Frob}
	\norm{\check\vZ^+\check \vX - \hat\vU^+}_F \le (1+\gee) \min_{\V{\gC} \in \sM_{n''\times K}, \vX \in \R^{K\times K}}  \norm{\V{\gC} \vX - \hat\vU^+}_F.\eeq
	Finally, $\check\vZ^+$ is extended to $\check\vZ\in\sM_{n',K}$, and then 
	$\check\vZ$ is extended to $\check\vZ_0\in\sM_{n,K}$ by taking $\check\vZ_{l_j,*}:=\check\vZ^+_{j,*}, j\in [n''],$ and $(\check\vZ_0)_{k_j,*}:=\check\vZ_{j,*}, j\in [n'],$  and filling in the remaining rows arbitrarily.
	$\check\vZ_0$ is the estimate of $\vZ$ from this method. Unlike in Algorithm 1, we use the normalized nonzero rows of $\hat\vU$ in  Algorithm 2 (see~\eqref{k mean_Frob}) instead of all rows of $\hat\vU$ in Algorithm 1 (see \eqref{eq:kmeans}). However, like in Algorithm 1, the reason for using an $(1+\gre)$-approximate $K$-means clustering algorithm in Algorithm 2 is also purely theoretical as we need theoretical guarantee on the solutions of the heuristic algorithms used to solve the $K$-means problem as given in works like \cite{kumar2004simple, feldman2007ptas}.   
	
	\vspace{0.2in}
	\framebox[\textwidth]
	%\label{algo_1}
	{\centering\parbox{.95\textwidth}{
			\textbf{Algorithm 2:} Spherical Spectral Clustering of the Sum of the Squared Adjacency Matrices \\
			\textbf{Input:} Adjacency matrices $\vA^{(1)}, \vA^{(2)}, \ldots, \vA^{(T)}$; number of communities $K$; approximation parameter $\gre$. \\ 
			%truncation parameter $\gd$.\\
			\textbf{Output:} Membership matrix $\check\vZ_0$. \\
			\begin{enumerate}
				\item Perform steps 1-7 of Algorithm 1.
				\item Let $n''$ be the number of nonzero rows (having indices $1\le l_1<l_2<\cdots <l_{n''}\le n'$) of $\hat\vU$. Obtain $\hat\vU^+ \in \R^{n''\times K}$ consisting of normalized nonzero rows of $\hat\vU$, i.e.~$\hat\vU^+_{i,*}=\hat\vU_{l_i,*}/\norm{\hat\vU_{l_i,*}}_2$ for $i\in[n'']$. 
				\item Use $(1+\gre)$ approximate $K$-means clustering algorithm on the row vectors of $\hat\vU^+$ to obtain $\check\vZ^+ \in \sM_{n'',K}$ and $\check \vX \in \R^{K \times K}$ obeying \eqref{k mean_Frob}.
				\item Extend $\check\vZ^+$ to obtain $\check\vZ\in\sM_{n',K}$ as follows. $\check\vZ_{j,*} = \check\vZ^+_{i,*}$ (resp.~$(1,0,\ldots, 0)$) for $j=l_i$ (resp.~$j\notin\{l_1, \ldots, l_{n''}\}$). 
				\item Extend $\check\vZ$ to obtain $\check\vZ_0\in\sM_{n,K}$ as follows. $(\check\vZ_0)_{j,*} = \check\vZ_{i,*}$ (resp.~$(1, 0,\ldots, 0)$) for $j=k_i$ (resp.~$j\notin\{k_1, \ldots, k_{n'}\}$). 
				\item $\check\vZ_0$ is the estimate of $\vZ$.
			\end{enumerate}
	}}\\
	
	\subsection{Selection of $K$}
	\label{sec_k_sel}
	In both Algorithm 1 in \textsection \ref{sec_algo_1} and Algorithm 2 in \textsection \ref{sec_algo_2}, the number of communities $K$ were considered to be known. However, number of communities can also be estimated using the absolute eigenvalues of the matrix $\vA^{[2]}$ by using the thresholding methods as in \cite{chatterjee2015matrix, bickel2016hypothesis, le2015estimating}. More work needs to be done to get a better estimate of number of communities $K$ in the multiple network context. Extensions of methods in \cite{bickel2016hypothesis}, \cite{wang2017likelihood}, \cite{chen2018network}, and \cite{le2015estimating} seem to be the first step for further~research~on~this~topic.
	
	Here, we give an intuitive method for detection of number of communities based on the eigenvalues of $\vA^{[2]}$. Using the concentration result of $\vA^{[2]}$ to $\E\vA^{[2]}$ used in proof of Theorem \ref{ConsSum1}, we can get a threshold on the eigenvalues of $\vA^{[2]}$ corresponding to the zero eigenvalues of $\vA_0^{[2]}$.
	
	\vspace{0.2in}
	\framebox[\textwidth]
	%\label{algo_1}
	{\centering\parbox{.95\textwidth}{
			\textbf{Algorithm 3:} Detecting Number of Communities using Sum of the Squared Adjacency Matrices \\
			\textbf{Input:} Adjacency matrices $\vA^{(1)}, \vA^{(2)}, \ldots, \vA^{(T)}$. \\
			\textbf{Output:} Estimated number of communities $\hat{K}$. \\
			\begin{enumerate}
				\item Perform steps 1-6 of Algorithm 1.
				\item Obtain $\gl_1\geq \gl_2 \geq \cdots \geq \gl_n$ as the eigenvalues of $\vA^{[2]}$.
				\item Define estimated number of communities as $\hat{K} = \argmin\left\{k: \gl_k >\frac 14 \left(T\bar{d^2}\right)\left(T\left[\bar{d^2}\right]^{1/2}\right)^{-1/8}\right\}$. 
			\end{enumerate}
	}}\\

	\section{Theoretical Justification}
	\label{sec_theory}
	
	\subsection{Consistency of  spectral clustering label $\hat\vZ_0$ under multilayer stochastic block model}
	\label{sec_proof_hatZ_supp}
	In order to state the theoretical results on the estimated community membership matrix, $\hat\vZ_0$, we first need to define certain quantities and conditions on the parameters of multilayer stochastic block model. 
	%Let us consider a multi-relational network $\left(\vA^{(1)}, \ldots, \vA^{(T)}\right)$ generated from the multilayer stochastic block model (MSBM) with parameters $(\vz, \V{\pi}, \vB)$. 
	The following parameters are functions of $(\vz, \V{\pi}, \{\vB^{(t)}\}_{t=1}^T)$: (i) {\it $d=n(\max_{a,b\in[K], t\in[T]}B^{(t)}_{ab})$ is the maximum expected degree} of a node at any layer; (ii) {\it $\gl=T^{-1} \sum_{t\in[T]}\gl_K\left((\frac nd \vB^{(t)})^2\right)>0$ is the average of the smallest eigenvalues} of squared normalized probability matrices $\{\vB^{(t)}\}_{t=1}^T$; and (iii)  $n_{\text{min}}$ is the size of the smallest community.
	
	\begin{thm} \label{ConsSum1}
		Let $(\vA^{(t)}, t\in[T])$ be the adjacency matrices of the networks generated from the multilayer stochastic block model with parameters $(\vz, \V{\pi}, \{\vB^{(t)}\}_{t=1}^T)$. For $a\in[K]$, let $f_a$ denote the proportion of nodes having community label $a$, which are misclassified in Algorithm 1. For any $\gee>0$ and $\gD>8$, there are constants $C=C(\gee),  C'>0$ such that if  
		\begin{align} \label{eq_sum_ass}
		% \frac 1T \sum_{t\in[T]}  \gl_K\left(\left[\vB^{(t)}\right]^2 \right) =: 
		\gl  \left(\frac{n_{\text{min}}}{n}\right)^2 >\max\left\{\frac 7n, \frac{C\gD\sqrt K}{ (Td)^{1/4}}\right\}, \text{ then }
		\end{align}
		\begin{align}
		\pr\left(\sum_{a\in[K]} f_a \le   \left[\frac{C\gD\sqrt K}{(Td)^{1/4} \gl  \left(\frac{n_{\text{min}}}{n}\right)^2}\right]^2\right) \ge  1- \frac{C'+2nK}{n(Td)^{3/4}} - 2n^{5-\gD^2/12}. \label{eq_mis_error}
		%\exp\left(-\left[\frac{\gD^2}{12}-5\right]\log(n)\right).
		\end{align}
		Therefore, in the special case,  when (i) $K$ is a constant and
		(ii) the community sizes are balanced, i.e.~$n_{\text{max}}/n_{\text{min}} = O(1)$, then the proportion of misclassified nodes in $\hat\vZ_0$ is arbitrarily small (resp.~goes to zero) with probability $1-o(1)$ if $(Td)^{1/4}\gl$ is large enough (resp.~$(Td)^{1/4}\gl\to\infty$).
	\end{thm}
	
	\begin{rem}\label{rem_condition_asymp}
		Note that the result in equation \eqref{eq_mis_error} involves the interplay of the parameters $n$, $T$, $K$ and $\{\vB^{(t)}\}_{t=1}^T$ and does not assume any apriori condition on any of the parameters except equation \eqref{eq_sum_ass}. Also, the result in equation \eqref{eq_mis_error} is a non-asymptotic result, but, it can be made into an asymptotic result. We need the condition 
		%$ \sqrt{K}\gD(Td)^{-1/4}\rightarrow 0$ and $\gD^2\log(n)\to\infty$ for having asymptotically vanishing error with probability $1-o(1)$. The asymptotics can be with respect to $T\to \infty$ and/or $d_n\to \infty$ as $n \to \infty$. 
		$ \sqrt{K}(Td)^{-1/4}\gl^{-1}\rightarrow 0$ and 
		%	$Td\to\infty$ 
		$\gl > \frac{cn}{n_{min}^2}$ for $c > 7$ for having an asymptotically vanishing mis-classification error with probability $1-o(1)$. The asymptotics can be with respect to $T\to \infty$ and/or $d\to \infty$ (as $n \to \infty$).
		%Both the constants $\gd>0$ and $\eta>0$ can be taken small. We can even get rid of $\eta$ and consider any increasing function of $Td^2$ instead of $(Td^2)^{2\eta}$ in equation \eqref{eq_mis_error}. 
		
		Also, for the asymptotic case of $n\to\infty$, $\gD$ is a constant. But for $n$ fixed and $T\to\infty$, $\gD$ has to be chosen such that it satisfies both $n^{5-\gD^2/12}\to 0$ and the condition in equation \eqref{eq_sum_ass}. For example $\gD$ can be taken as, $\gD = \left(\frac{(Td)^{1/4} n_{min}^2}{\sqrt{K}n^2}\right)^{\gd}$ for any constant $\gd$ with $0 < \gd < 1$.
	\end{rem}
	
	\begin{rem}\label{rem_condition_optimal}
		The condition ``$(Td)^{1/4}\gl\to\infty$" for the special case, is necessary and sufficient in order to have a consistent estimator of $\vZ$.
		Theorem \ref{ConsSum1} proves the sufficiency. The  necessity of the condition follows from the work of \cite{ZZ16}. Consider a stochastic block model (so $T=1$), where (i) there are two communities having equal size $n$ and  (ii) the within (resp.~between)  community connection probability is $a/n$ (resp.~$b/n$) for some constants $a>b>0$. In this case $(Td)^{1/4}\gl=\frac{a^{1/4}(a-b)}{a}$  is a constant.  \cite{ZZ16} states that in the above setup, there is a constant $c>0$  such that if 
		\[ \frac{(a-b)^2}{a+b} <c\log\frac 1\gc\]
		for some constant $\gc$ (e.g.~$\gc=e^{-(a-b)/c}$), 
		then the expected proportion of misclassification for every algorithm will be at least $\gc$. In other words, no algorithm can give consistent estimator~of~$\vZ$. So, the condition ``$(Td)^{1/4}\gl\to\infty$" becomes an \emph{optimal} condition for consistent recovery of community labels.
	\end{rem}
	
	\begin{rem}\label{rem_condition_eig}
		The assumption in equation \eqref{eq_sum_ass} makes sure that there is a community structure in the aggregated network. The condition in \eqref{eq_sum_ass} is quite relaxed. In the balanced case with constant $K$, it does not require $O(T)$ many matrices among $\{\vB^{(t)}\}_{t=1}^T$ to have full-rank but only requires $\frac{T}{(Td)^{1/4}}$, which is $o(T)$, many matrices among $\{\vB^{(t)}\}_{t=1}^T$ to have all nonzero eigenvalues or full-rank. Note that according to the condition in equation \eqref{eq_sum_ass}, the number of necessary informative (full-rank) $\{\vB^{(t)}\}_{t=1}^T$ matrices (a) should increase as $T$ increases for fixed but large $d$ (or $n$); (b) should decrease as $d$ (or $n$) increases for fixed $T$. This behavior is illustrated in Scenario 2 of simulation in \textsection \ref{sec_simulation}.
	\end{rem}
	
	\subsection{Extensions to the case of dependent adjacency matrices} \label{sec:DepAdj}
	In this section, we will consider a general situation where $\cB := (\vB^{(t)}, t\in[T])$ is a stochastic process and the distribution of $(\vA^{(t)}, t\in[T])$ is conditionally independent as described in \eqref{A^t bmdef}. Now, let us define some important functions of the stochastic process $(\vB^{(t)}, t\in[T])$ which will be useful in quantifying the mis-classification error of Algorithm 1. 
	\begin{enumerate}[(a)]
		\item The smallest eigenvalue $\gl_{K,t}:=\gl_K([\vB^{(t)}]^2)$ is also a random variable and cumulative distribution function of $\gl_{K,t}$ is given by $F_t(x):=\pr(\gl_{K,t}\le x)$ for $x\ge 0$. Let $b_t:=\mathbf 1_{\{\gl_{K,t}=0\}}$ be the indicator random variable for the event of rank-deficient $[\vB^{(t)}]^2$. Let $F^+_t(x):=\frac{F_t(x)-F_t(0)}{1-F_t(0)}$ be the distribution function corresponding to the truncated positive part of $\gl_{K,t}$ and $\tilde\gl_{K,t}\sim F^+_t(x)$ for all $t$ is an independent copy generated from the truncated distribution. So, $\tilde\gl_{K,t}$ is independent of $\cB$. Then, we can define the random variable
		\[\gl^+_{K,t}:=\begin{cases}\gl_{K,t} \text{ if } b_t=0, \\\tilde\gl_{K,t}\text{ if } b_t=1.\end{cases}\] 
		So $\gl_{K,t}= b_t\Gd_0+(1-b_t)\gl^+_{K,t}$, $b_t\sim Ber(F_t(0)), \gl^+_{K,t}\sim F^+_t$. Lastly, it follows from elementary probability calculations that $\mathfrak b=(b_t, t\in[T])$ and $\boldsymbol\gl=(\gl^+_{K,t}, t\in[T])$ are independent. 
		
		\item The \emph{maximal degree variable},  $\ul d_n(\eps)$ for any $\eps > 0$, is defined in the following way -
		\begin{align*}
		\ul d_n(\eps):=\sup\left\{x\in[0, n]: \pr\left(\max_{t\in[T], \; a,b\in[K]} n B^{(t)}_{ab}\le x\right)\le \eps\right\}.
		\end{align*}
		
		\item \textbf{(Mixing condition)} We consider a decreasing function $\ga_\da:\Z_+\mapsto[0,1]$ to reflect the decay of correlation (at any rate) between two events of non-informative (smallest eigenvalue of $\vB^{(t)}$ being zero) $\vB^{(t)}$ matrices, like $\vB^{(t_1)}$ and $\vB^{(t_2)}$, where, $ t_1,t_2 \in [T], t_1\neq t_2$.
		\begin{align}
		\left|\pr\left(\cap_{i\in[2]}\{\gl_{K,t_i}=0\}\right) - \prod_{i\in[2]}\pr(\gl_{K,t_i}=0)\right| \le \ga_\da(|t_1-t_2|) \label{StMix}
		\end{align}
		with $\ga_\da$ having the property
		\[\ga_\da(s) \da 0 \text{ as $ s\ua \infty$, and } \ga_\da(0) = 1. \]
		This decay of correlation is necessary to have consistent recovery of communities.
		
		\item We consider a function $\psi_{\ua\da}:\N\times\R_+\mapsto[0, 1]$ in terms of $T$ and $\ul d_n(\eps)$, which captures the probability that network layers are  non-informative, that is,
		\begin{align}
		\max_{t \in [T]}\pr\left(\{\gl_{K,t} = 0\}\right) \leq \psi_{\ua\da}(T, \ul d_n(\eps)) . \label{ind_net} 
		\end{align}
		$\psi_{\ua\da}(T, \ul d)$ is a function which captures the behavior that on one hand $\psi_{\ua\da}$ increases to 1 as $T$ increases and $\ul d$ remains constant. But, on the other hand $\psi_{\ua\da}$ decreases to 0 as $\ul d$ increases and the number of networks $T$ stays the same, that is, 
		\[ \lim_{T\ua\infty}\psi_{\ua\da}(T,\ul d) = 1\ \ \ \text{and }\ \ \psi_{\ua\da}(T,\ul d) \da 0 \text{ as $ \ul d\ua \infty$. }\] 
		%	An example of such a function can be $\psi_{\ua\da}(T,\ul d) = (1 - 1/(T\ul d))\wedge (1/\ul d)$.
		
		\item We consider a decreasing and convex function $\phi_\da:(0,\infty)\mapsto(0,\infty)$, which controls the lower tail behavior of the smallest eigenvalues of the probability matrices $(\vB^{(t)}, t\in[T])$ near 0 with the property
		\[\phi_\da(x)\ua\infty \text{ as $ x\da 0$. }\]
	\end{enumerate}
	Based on the random variables $\gl_{K,t}^+$ and $\ul d_n(\eps)$, and the functions $\ga_\da$, $\psi_{\ua\da}$ and $\phi_\da$ defined above, we place the following conditions on the stochastic process $\cB$. \\
	\textbf{Assumption A: } Let $\cB = (\vB^{(t)}, t\in[T])$ be a stochastic process with the following properties - 
	\begin{eqnarray}
	& & (a)\; \psi_{\ua\da}(T, \ul d_n(\eps))\le 1-\left[\frac{\sqrt T}{T}+\ga_\da(\sqrt T)\right]^{1/2-\gd}\vee\frac{1}{(T\ul d_n(\eps))^{\frac{1}{60}}} \notag \\
	& & (b)\; \max_{t\in[T]} F_t(0)\le \psi_{\ua\da}(T,\ul d_n(1/2)), \text{and }  \label{lamAss} \\
	& & (c)\; \max_{t\in[T]}\E\phi_\da(\gl^+_{K,t}) \le C_1 \notag
	\end{eqnarray} 
	for any $\eps>0$ and for some constants $C_1<\infty \text{ and } \gd<1/2$.

	\begin{thm} \label{Gen B^t}
		Let $(\vB^{(t)},t\in[T])$ be any stochastic process satisfying Assumption A of \eqref{lamAss}, $(\vA^{(t)}, t\in[T])$ be the adjacency matrices satisfying \eqref{A^t bmdef}, $(f_a, a\in[K])$ and $C'$ be as in Theorem \ref{ConsSum1}. For any $\eps>0, \gd\in(0,1/2)$,
		\begin{align} 
		\pr\left(\sum_{a\in[K]} f_a >   (T\ul d(\eps))^{-1/6}\right) \le  \eps+  \frac{C_1}{\phi_\da\left(\frac{2n^2}{n^2_{\text{min}}}\left[T\ul d_n(\eps)\right]^{-1/15}\right)} \label{eq_mis_error_mart}\\
		+\min\left\{4\left[\frac{\sqrt T}{T}+\ga_\da(\sqrt T)\right]^{2\gd}, T\psi_{\ua\da}(T,\ul d_n(\eps))\right\} +\frac{2C'+2nK}{n}\left([T\ul d_n(\eps)]^{-3/4}+\eps\right). \notag
		\end{align}
		Therefore, in the special case,  when (i) $K$ is a constant and
		(ii) the community sizes are balanced, i.e.~$n_{\text{max}}/n_{\text{min}} = O(1)$, then the proportion of misclassified nodes in $\hat\vZ_0$ is arbitrarily small (resp.~goes to zero) with probability $1-o(1)$ if $T\ul d_n(\eps)$ is large enough (resp.~$T\ul d_n(\eps)\to\infty$) and $\eps $ is small enough (resp.~$\eps\to 0$).
	\end{thm}
	\begin{rem}
		The statement of Theorem \ref{Gen B^t} is pretty general, flexible and involves many components which can be fine-tuned to capture a wide-variety of aspects of the community detection problem under dependence between layers.
		\vspace{0.04in}
		% \begin{enumerate}
		
		1. The assumption on $\ga_\da$ reflects the decay of correlation (at any rate) between the two events of non-informative (smallest eigenvalue of $\vB^{(t)}$ being zero) $\vB^{(t)}$ matrices, like $\vB^{(t_1)}$ and $\vB^{(t_2)}$, where, $ t_1,t_2 \in [T], t_1\neq t_2$. This decay of correlation is necessary to have consistent recovery of communities. Faster decay rate implies smaller error rate in terms of $T$ for community recovery.
		\vspace{0.04in}
		
		2. The assumption about $\psi_{\ua\da}$ shows that consistent recovery of communities is possible  by bounding the probability that an individual network is non-informative (i.e.~corresponding $\vB^{(t)}$ is singular).
		%(i) increases to 1 (at a certain maximum rate) as $T$ increases and sparsity of individual networks stay the same, or (ii) decreases to 0 (at any rate) as  sparsity of individual networks increases and the number of networks stays the same. The later condition is also necessary to have consistent recovery of communities.
		\vspace{0.04in}
		
		3. The assumption on $\phi_\da$ describes the lower tail behavior of the smallest eigenvalues of the probability matrices $(\vB^{(t)}, t\in[T])$ near 0. The larger value of $\phi_\da$ implies smallest eigenvalue being further away from zero and thus smaller error rate.
		\vspace{0.04in}
		
		4. The reason for taking $\ul d_n(\eps)$ as a measure of sparsity is the following.  When $T$ is small, in order to have a consistent algorithm it is necessary for not only the mean, but also all quantiles of the distribution of the maximum degree to increase to infinity.  If $T$ is small and the distribution of $\max_{t,a,b} B^{(t)}_{ab}$ has non-vanishing probability  for any  subinterval of $\R_+$, then consistency cannot hold. Note that, if $d^{(T)} := \max_{t\in[T], \; a,b\in[K]} n B^{(t)}_{ab}$ concentrates, then, $\E(d^{(T)})$ or $\text{Median}(d^{(T)})$ can replace $\ul d_n(\eps)$ in the Theorem.
		\vspace{0.04in}
		
		5. The $\sqrt T$ appearing in \eqref{eq_mis_error_mart} can be replaced by any $o(T)$ term.
		\vspace{0.01in}
		
		6. The exponents $1/6$ and $1/15$ appearing in \eqref{eq_mis_error_mart} vary between 0 and $1/2$. If $1/6$ is replaced by $\eta$, then $1/15$ can be replaced by any number $<1/4-\eta/2$. 
		
		7. The asymptotics in Theorem \ref{Gen B^t} can be with respect to $T\to \infty$ and/or $d\to \infty$ (as $n \to \infty$). The rates of decay of functions $\ga_\da$, $\psi_{\ua\da}$ and $\phi_\da$ also become crucial for mis-classification error to vanish with probability $1-o(1)$ as $T\to \infty$ and/or $d\to \infty$ (as $n \to \infty$).
		%Both the constants $\gd>0$ and $\eta>0$ can be taken small. We can even get rid of $\eta$ and consider any increasing function of $Td^2$ instead of $(Td^2)^{2\eta}$ in equation \eqref{eq_mis_error}. 
	\end{rem}
	
	\begin{cor}\label{cor_Gen_Bt}
		If $n$ is constant and $(\vB^{(t)}, t\in[T])$ is jointly stationary and ergodic process with $\gl:=\E\gl_K([\vB^{(1)}]^2)>0$, then there is a sequence of fractions $\{\eps_T\}_{T\ge 1}$ satisfying $\eps_T\da 0$ as $T\ua \infty$ such that $\pr(\sum_{a\in[K]} f_a<cKT^{-1/4}\gl^{-2})\ge 1-(c_1+c_2K)T^{-3/4}-\eps_T$, where $c, c_1, c_2$ are constants. 	
	\end{cor}
	
	\begin{rem}
		In the setup of Theorem \ref{ConsSum1} and \ref{Gen B^t},  if $\{\vB^{(t)}\}_{t=1}^T$ is a piecewise constant stochastic process with $k(T)$ many change-points, and the adjacency matrices  remain unchanged between successive change-points and we apply Algorithm 1 on the distinct adjacency matrices, then all the communities can be recovered consistently if $(k(T)d)^{1/4}\gl\to \infty$ (resp.~$(k(T)\ul d_n(\eps))^{1/4}(\min_{t\in[k(T)]}\gl_{K,t})\to \infty$) in the case of Theorem \ref{ConsSum1} (resp.~\ref{Gen B^t}).
	\end{rem}
	
	\subsection{Consistency of Spherical Spectral Clustering Labels $\check\vZ_0$ under multilayer degree-corrected block model}
	\label{sec_mdsbm}
	In order to state the theoretical results on the estimated community membership matrix, $\check\vZ_0$, we first need to define certain quantities and conditions on the parameters of multilayer degree-corrected block model. 
	%Let  $\left(\vA^{(1)}, \ldots, \vA^{(T)}\right)$ be the adjacency matrices of a sequence of networks generated from the multiple degree-corrected block model with parameters $(\vz, \V{\pi}, \{\vB^{(t)}\}_{t=1}^T, \V{\psi})$. 
	The following parameters are functions of $(\vz, \V{\pi}, \{\vB^{(t)}\}_{t=1}^T, \V{\psi})$: (i) $d=n(\max_{a,b\in[K], t\in[T]}B^{(t)}_{ab}))$ is the maximum expected degree of a node at any snapshot; (ii) \; for $a\in[K]$, $\tilde n_a:=\sum_{i\in\cC_a\}}\psi_i^2$ and $\tau_a := \sum_{i\in\cC_a} \psi_i^2 \sum_{i\in\cC_a} \psi_i^{-2}$ is a measure of heterogeneity of $\V{\psi}$; (iii) $\psi_{\text{min}}:=\min_{i\in[n]} \psi_i$; (iv) $\tilde n_{\text{max}}=\max_{a\in[K]}\tilde n_a, \tilde n_{\text{min}}=\min_{a\in[K]}\tilde n_a$; and (v) $\gl=T^{-1} \sum_{t\in[T]}\gl_K\left((\frac nd \vB^{(t)})^2\right)>0$ the  average  of the smallest eigenvalues of the squared normalized probability matrices. 
	\begin{thm} \label{ConsSum2}
		Let $(\vA^{(t)}, t\in[T])$ be the adjacency matrices of the networks (having $n$ nodes and $K$ communities) generated from the multilayer degree-corrected block model with parameters $(\vz, \V{\pi}, \{\vB^{(t)}\}_{t=1}^T, \V{\psi})$ satisfying \eqref{eq_dcbm_id}.  For any $\gee>0$ and $\gD>8$, there are constants $C(\gee), C'>0$~such~that~if 
		\begin{align} \label{lambda_assump_alt}
		% \frac 1T \sum_{t\in[T]}  \gl_K\left(\left[\vB^{(t)}\right]^2 \right) =: \gl >\frac{(n/\tilde n_{\text{min}})^2}{\min\{\frac n7, \frac 12 (Td)^{\gf/2}\}},
		\gl  \left(\frac{\tilde n_{\text{min}}}{n}\right)^2 >\frac 7n,\text{ and } 
		n_{\text{min}}> \frac{C(K\tilde n_{\text{max}})^3\gl^{-2}}{\psi^2_{\text{min}}\tilde n^4_{\text{min}}} + \frac{C\gD\sqrt{K\sum_{a\in[K]}\tau_a}}{ (Td)^{1/4}\gl  \left(\frac{\tilde n_{\text{min}}}{n}\right)^2},\end{align}
		then the total number  of misclassified nodes in $\check\vZ_0$ is at most
		\beq \label{overall misclassify} 
		\frac{C(K\tilde n_{\text{max}})^3}{(\psi_{\text{min}}\gl)^2(\tilde n_{\text{min}})^4} + \frac{n+C\gD\left(K\sum_{k\in[K]} \tau_k\right)^{1/2}}{ (Td)^{1/4}\gl\left(\frac{\tilde n_{\text{min}}}{n}\right)^2}\eeq
		with probability at least $1-(C'/n+2K)(Td)^{-3/4}-2n^{5-\gD^2/12}$.
		
		Therefore, in the special case,  when (i) $K$ is a constant,
		(ii) the community sizes are balanced, i.e.~$n_{\text{max}}/n_{\text{min}} = O(1)$ and (iii) $\psi_i=\alpha_i/\max\{\alpha_j: z_i=z_j\}$, where $(\alpha_i)_{i=1}^n$ are  i.i.d.~positive weights, then consistency holds for $\check\vZ_0$ with probability $1-o(1)$ if $\E[\max\{\alpha_1^2,\alpha_1^{-2}\}]<\infty$ and $(Td)^{1/4}\gl\to\infty$.
	\end{thm}
	
	\begin{rem}
		The condition ``$(Td)^{1/4}\gl\to\infty$" for the special case, is necessary and sufficient in order to have a consistent estimator of $\vZ$.
		Theorem \ref{ConsSum2} proves the sufficiency. The necessity of the condition also follows from the work of \cite{ZZ16} by considering the special case of stochastic block model. So, the condition ``$(Td)^{1/4}\gl\to\infty$" becomes an \emph{optimal} condition for consistent recovery of community labels.
	\end{rem}
	\begin{rem}
		Like in Remark \ref{rem_condition_asymp}, the statement in Theorem \ref{ConsSum2} is also non-asymptotic, but it can be viewed as an asymptotic statement in terms of $T \to \infty$ and/or $d\to\infty$ (as $n \to \infty$) under conditions \eqref{lambda_assump_alt} and \eqref{eq_dcbm_id}.
	\end{rem}
	\begin{rem}
		In the special case for $(\alpha_i)_{i=1}^n$ as  i.i.d.~positive weights, condition of $\E[\max\{\alpha_1^2,\alpha_1^{-2}\}]<\infty$ is satisfied for a large class of distributions, such as $\text{Uniform}(c, d)$ with $c > 0$, $\text{Pareto}(\ga)$ with $\ga > 2$ and $\text{Gamma}(\ga, \gb)$ with $\ga > 2$. 
	\end{rem}

	\subsection{Consistency of Estimated Number of Communities $\hat{K}$ in Algorithm 3}
	\label{sec_Khat_th}
	
	In \textsection \ref{sec_k_sel}, we give a method for estimating number of communities in Algorithm 3. In order to prove consistency of the estimated number of communities, $\hat{K}$, obtained from Algorithm 3, we consider that the multi-relational network has been generated from the multiple stochastic block model with parameters $(\vz, \V{\pi}, \{\vB^{(t)}\}_{t=1}^T)$. 
	
	\begin{thm} \label{Khat_thm}
		Let $(\vA^{(t)}, t\in[T])$ be the adjacency matrices of the networks (having $n$ nodes and $K$ communities) generated from the multilayer stochastic block model with parameters $(\vz, \V{\pi}, \{\vB^{(t)}\}_{t=1}^T)$ and $\hat{K}$ be the estimate of  $K$ from Algorithm 3. Let us consider the special case when $K$ is a constant. There are constants 
		$C, C'>0$ such that if  $\gD>8, \gl(n_{\text{min}}/n)^2>\max\{7/n, 3(Td)^{-1/8}\}$ and $Td > C$,  then
		%\beq \label{Khat_mis} 
		$\pr\left(\hat{K} \neq K\right) \leq \frac{C'/n+2K}{(Td)^{3/4}} + 2n^{5-\gD^2/12}$.
		% \eeq
	\end{thm}
	
	\section{Simulation Results}
	\label{sec_simulation}
	We simulate multilayer networks in several different scenarios for empirically testing the community detection performance of the methods proposed in the paper. 
	
	We compare six different algorithms - 
	\begin{enumerate}[(i)]
		\item \emph{Sum:} spectral clustering with sum of adjacency matrices with truncation for high-degree nodes \cite{bhattacharyya2018spectral, bhattacharyya2020consistent}.
		\item  \emph{Spectral sum:} clustering the rows of sum of eigen-spaces $\sum_{t=1}^T U^{(t)}$ of each network snapshot (where, $\vU^{(t)}_{n\times K}$ is the matrix formed by the eigenvectors of top $K$ eigenvectors of $\vA^{(t)}$). It was shown empirically in \cite{paul2017spectral} to have a good community detection performance.
		\item \emph{Sum (Spherical):} spherical spectral clustering with sum of adjacency matrices with truncation for high-degree nodes \cite{bhattacharyya2018spectral, bhattacharyya2020consistent}.
		\item \emph{Co-regularized spectral clustering:} the method was proposed in \cite{kumar2010co} and shown empirically in \cite{paul2017spectral} to have a good community detection performance.
		\item \emph{Algorithm 1} of the paper.
		\item \emph{Algorithm 2} of the paper.
	\end{enumerate}	
	Note that all the algorithms are not compared in every experiment of the scenarios. We consider three different scenarios with different combinations of $n, T, \V{\psi}$, and $\vB$ to generate multilayer stochastic block models and multilayer degree-corrected block models. 
%	Another scenario based on assortative layers is presented in the Appendix \cite{bhattacharyya2020supplement}. 
	The performance on community detection is reported in terms of normalized mutual information (NMI) metric between true and estimated community labels. The value of NMI is between 0 and 1 and higher value of NMI implies better community detection performance.
	\vspace{0.1in}
	
	\noindent\emph{Scenario 1:} In this scenario, we consider a situation where the interaction between some communities change their nature from disassortative to assortative between layers where as interaction between some communities remain assortative throughout all the layers. We simulate such multilayer networks from multilayer stochastic block model (SBM) and multilayer degree-corrected block model (DCBM) under the framework of \eqref{eq_sbm1} and \eqref{eq_dcbm1} of \textsection \ref{sec_model}. We consider four experiments under this scenario. Each experiment is repeated 25 times and the results are averages over the repetitions.
	\vspace{0.04in}
	
\begin{figure}[htp]
		\begin{center}
			\subfigure[]{\includegraphics[height=4.5cm,width=6cm]{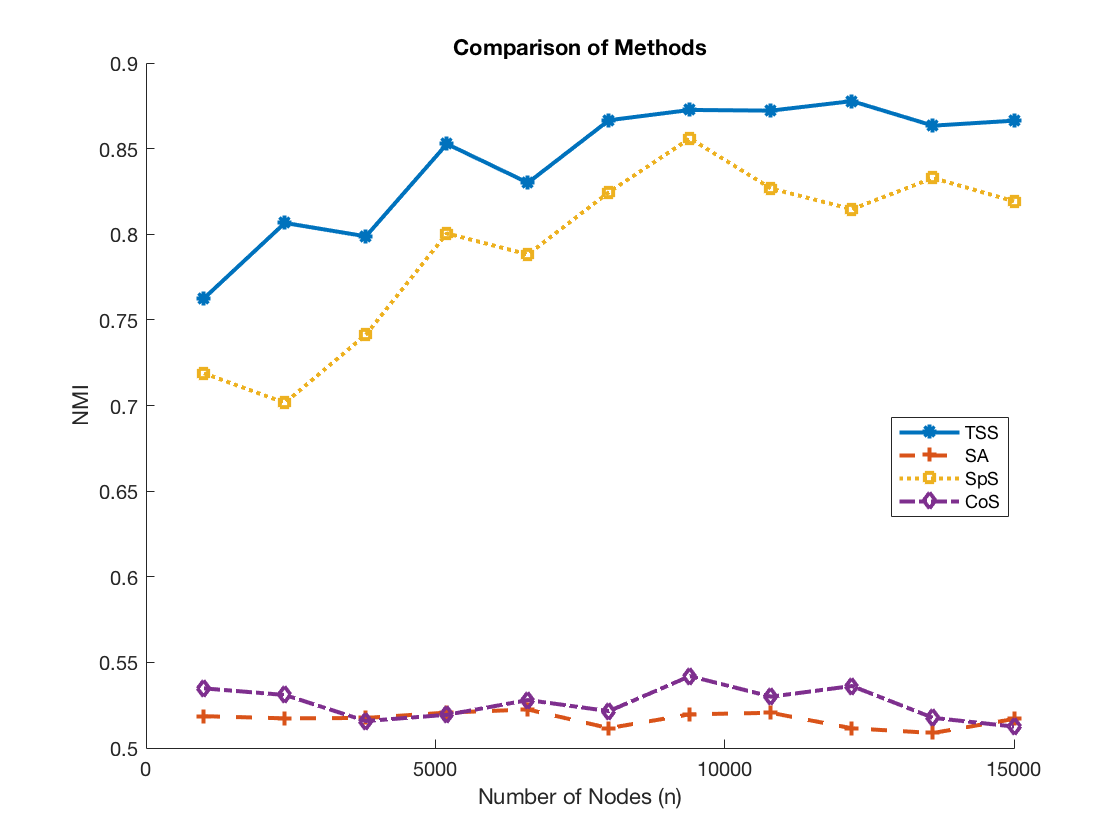}}
			\subfigure[]{\includegraphics[height=4.5cm,width=6cm]{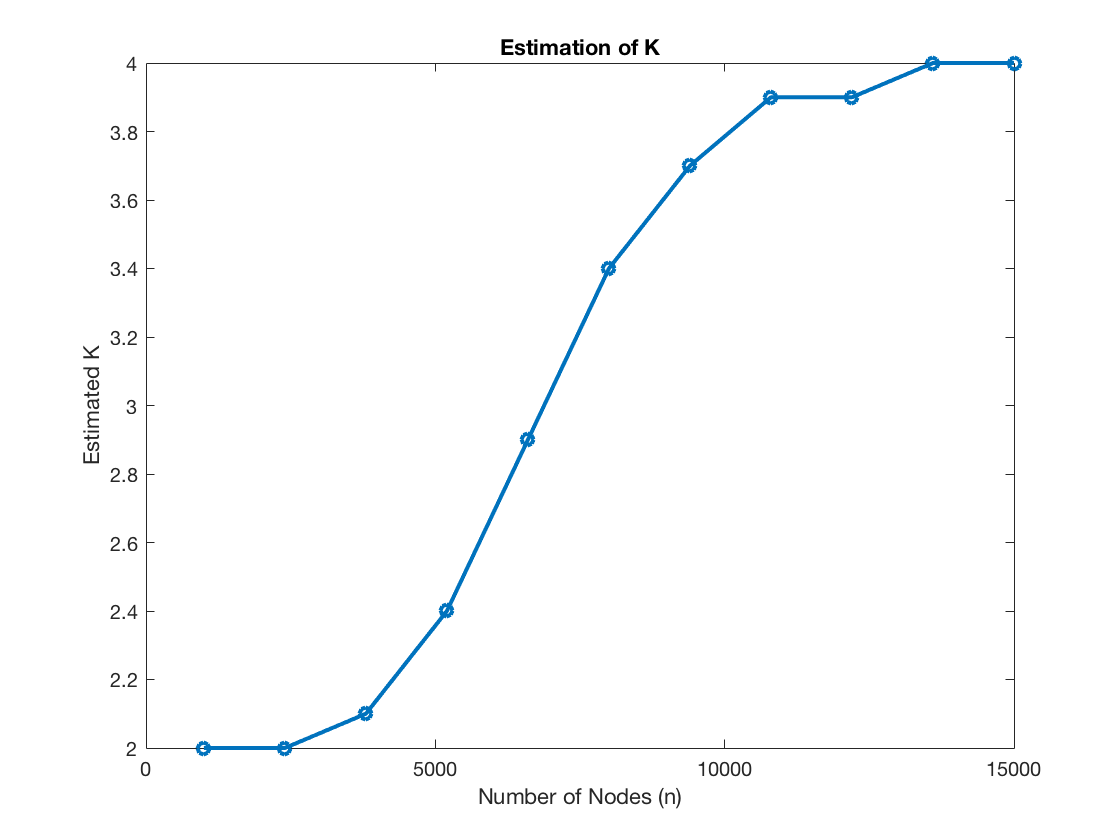}}
		\end{center}
		\caption{(a) NMI comparison using algorithms (i) as SA, (ii) as SpS, (iv) as CoS, and (v) as TSS; (b) Estimation of $K$ using Algorithm 3. }
		\label{fig_msbm_disass}
	\end{figure}

	%\begin{enumerate}
	1. \emph{Changing number of nodes ($n$) for multilayer SBM:} We vary node size $n$ from $1000$ to $15000$ with other parameters being $K=4$, $\vB^{(t)}_{4\times 4} = 3\frac{(\log n)^{3/4}}{n}\left(\V{I}_2\otimes\V{J}_2 + \ul b_t\vI_4\right)$, $\V{\pi} = \frac{1}{4}\V{1}_{4\times 1}$, and $T=11$, where, $\ul b_t = -1 + 0.2(t-1)$ for $t \in [T]$. We compare the algorithms (i), (ii), (iv), and (v). We also apply Algorithm 3 to estimate the number of communities with changing $n$. The results on average NMI and average $\hat{K}$ are given in Figure \ref{fig_msbm_disass}.
	\vspace{0.04in}
	
\begin{figure}[htp]
		\begin{center}
			\subfigure[]{\includegraphics[height=4.5cm,width=6cm]{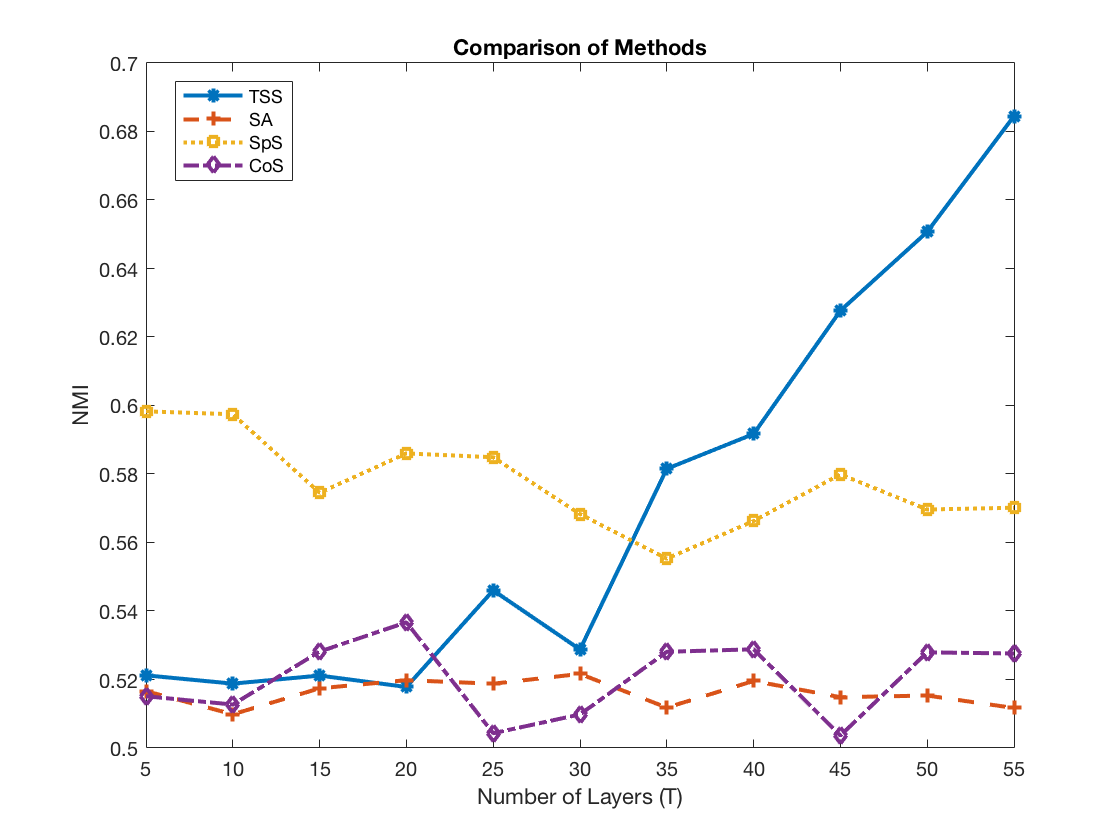}}
			\subfigure[]{\includegraphics[height=4.5cm,width=6cm]{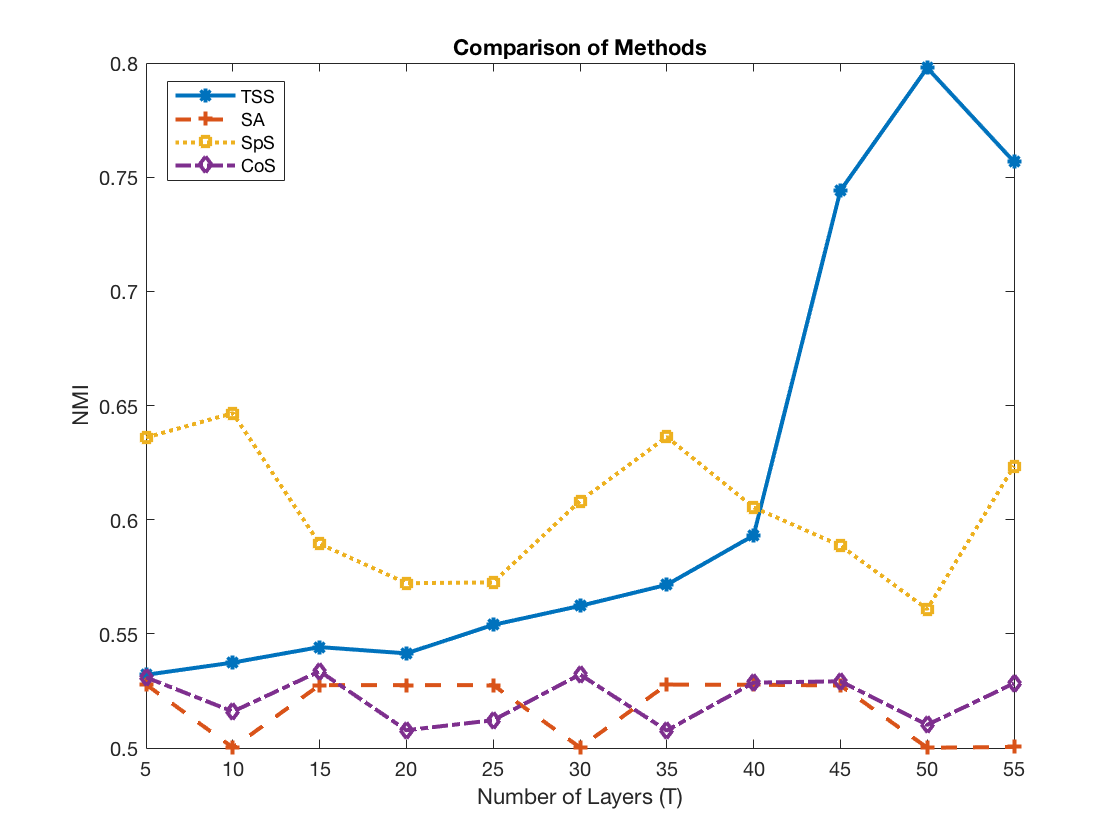}}
		\end{center}
		\caption{(a) NMI comparison using algorithms (i) as SA, (ii) as SpS, (iv) as CoS, and (v) as TSS; (b) NMI comparison using algorithms (ii) as SpS, (iii) as SA, (iv) as CoS, and (vi) as TSS. }
		\label{fig_msbm_mdcbm_disassT}
	\end{figure}

	2. \emph{Changing number of layers ($T$) for multilayer SBM:} We vary number of layers $T$ from $5$ to $55$ with other parameters being $K=4$, $\vB^{(t)}_{4\times 4} = \frac{5}{n}\left(\V{I}_2\otimes\V{J}_2 + \ul b_t\vI_4\right)$, $\V{\pi} = \frac{1}{4}\V{1}_{4\times 1}$, and $n=2000$, where, $\ul b_t = -1 + 0.2(t-1)$ for $t \in [T]$. We compare the algorithms (i), (ii), (iv), and (v). The results on average NMI are given in Figure \ref{fig_msbm_mdcbm_disassT}(a).
	\vspace{0.04in}
	
\begin{figure}[htp]
		\begin{center}
			\subfigure[]{\includegraphics[height=4.5cm,width=6cm]{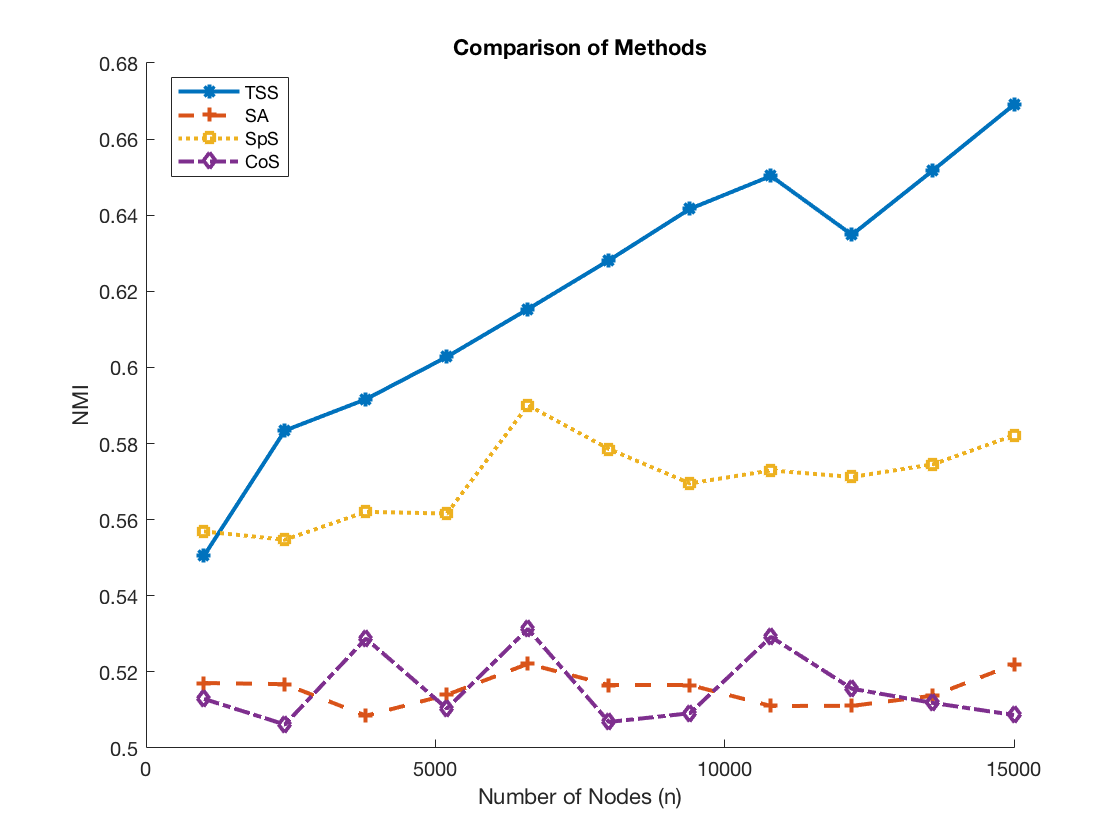}}
			%			\subfigure[]{\includegraphics[height=4cm,width=4cm]{plot_mdcbm_disass_KN1.png}}
		\end{center}
		\caption{(a) NMI comparison using algorithms (ii) as SpS, (iii) as SA, (iv) as CoS, and (vi) as TSS. }
		\label{fig_mdcbm_disass}
	\end{figure}

	3. \emph{Changing number of nodes ($n$) for multilayer DCBM:} We vary node size $n$ from $1000$ to $15000$ with other parameters remaining the same as experiment 1 with the only addition of degree 
	%being $K=4$, $\vB^{(t)}_{4\times 4} = 3\frac{(\log n)^{3/4}}{n}\left(\V{I}_2\otimes\V{J}_2 + \ul b_t\vI_4\right)$, $\V{\pi} = \frac{1}{4}\V{1}_{4\times 1}$, $\psi_i \stackrel{iid}{\sim} U(0.5, 1)$ for $i \in [n]$, and $T=11$, where, $\ul b_t = -1 + 0.2(t-1)$ for $t \in [T]$. 
	parameters $\psi_i \stackrel{iid}{\sim} U(0.5, 1)$ for $i \in [n]$. We compare the algorithms (ii), (iii), (iv), and (vi). The results on average NMI are given in Figure \ref{fig_mdcbm_disass}.
	\vspace{0.04in}
	
	4. \emph{Changing number of layers ($T$) for multilayer DCBM:} We vary number of layers $T$ from $5$ to $55$ with other parameters remaining the same as experiment 2 except $\vB^{(t)}_{4\times 4} = \frac{10}{n}\left(\V{I}_2\otimes\V{J}_2 + \ul b_t\vI_4\right)$ and the addition of degree
	%being $K=4$, $\vB^{(t)}_{4\times 4} = \frac{10}{n}\left(\V{I}_2\otimes\V{J}_2 + \ul b_t\vI_4\right)$, $\V{\pi} = \frac{1}{4}\V{1}_{4\times 1}$, $\psi_i \stackrel{iid}{\sim} U(0.5, 1)$ for $i \in [n]$, and $n=2000$, where, $\ul b_t = -1 + 0.2(t-1)$ for $t \in [T]$. 
	parameters $\psi_i \stackrel{iid}{\sim} U(0.5, 1)$ for $i \in [n]$. We compare the algorithms (ii), (iii), (iv), and (vi). The results on average NMI are given in Figure \ref{fig_msbm_mdcbm_disassT}(b).
	%\end{enumerate} 
	\vspace{0.1in}
	
	\noindent\emph{Scenario 2:} In this scenario, we consider a situation where only one layer has a disassortative community structure, where as all other network layers are uninformative in terms of the community structure. We simulate such multilayer networks under the framework of \eqref{eq_sbm1} and \eqref{eq_dcbm1} of \textsection \ref{sec_model}. We consider four experiments under this scenario. Each experiment is repeated 25 times and the results are averages over the repetitions.
	\vspace{0.04in}
	
\begin{figure}[htp]
		\begin{center}
			\subfigure[]{\includegraphics[height=4.5cm,width=6cm]{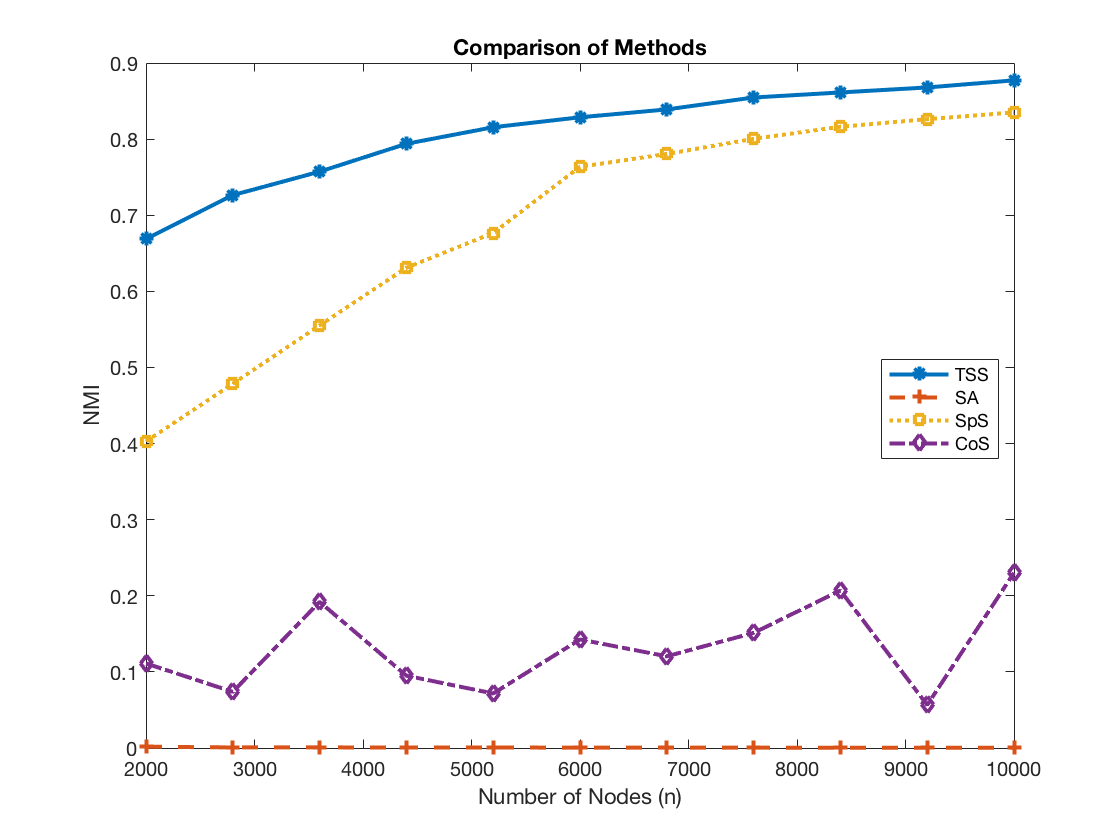}}
			\subfigure[]{\includegraphics[height=4.5cm,width=6cm]{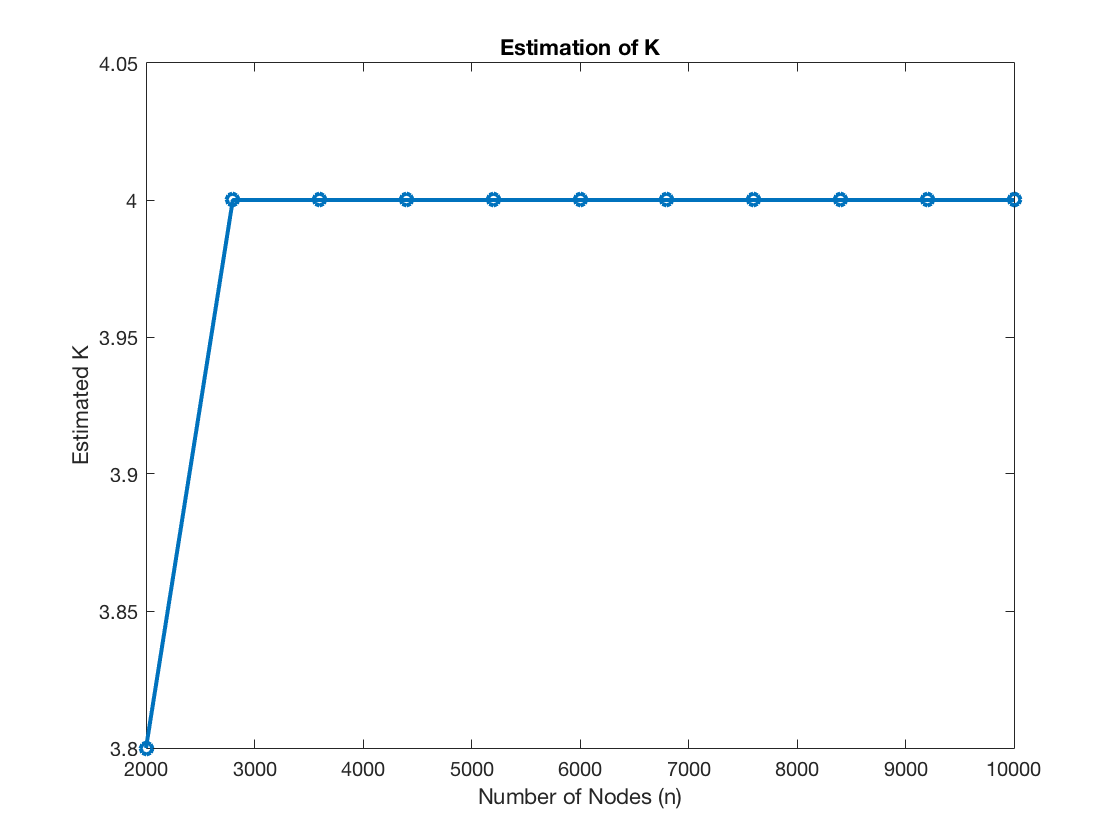}}
		\end{center}
		\caption{(a) NMI comparison using algorithms (i) as SA, (ii) as SpS, (iv) as CoS, and (v) as TSS; (b) Estimation of $K$ using Algorithm 3. }
		\label{fig_msbm_dis}
	\end{figure}

	%\begin{enumerate}
	1. \emph{Changing number of nodes ($n$) for multilayer SBM:} We vary node size $n$ from $2000$ to $10000$ with other parameters being $K=4$, $\vB^{(1)}_{4\times 4} = \frac{(\log n)^{4/3}}{n}\left(\V{J}_4 - \V{I}_4 \right)$, $\vB^{(2)}_{4\times 4} = \frac{(\log n)^{4/3}}{n}\V{J}_4$, $\vB^{(t)}_{4\times 4} = \frac{(\log n)^{4/3}}{nT}\V{J}_4 $ for $t=3, \ldots, T$, $\V{\pi} = \frac{1}{4}\V{1}_{4\times 1}$, and $T=11$. We compare the algorithms (i), (ii), (iv), and (v). We also apply Algorithm 3 to estimate the number of communities with changing $n$. The results on average NMI and average $\hat{K}$ are given in Figure \ref{fig_msbm_dis}.
	\vspace{0.04in}
	
\begin{figure}[htp]
		\begin{center}
			\subfigure[]{\includegraphics[height=4.5cm,width=6cm]{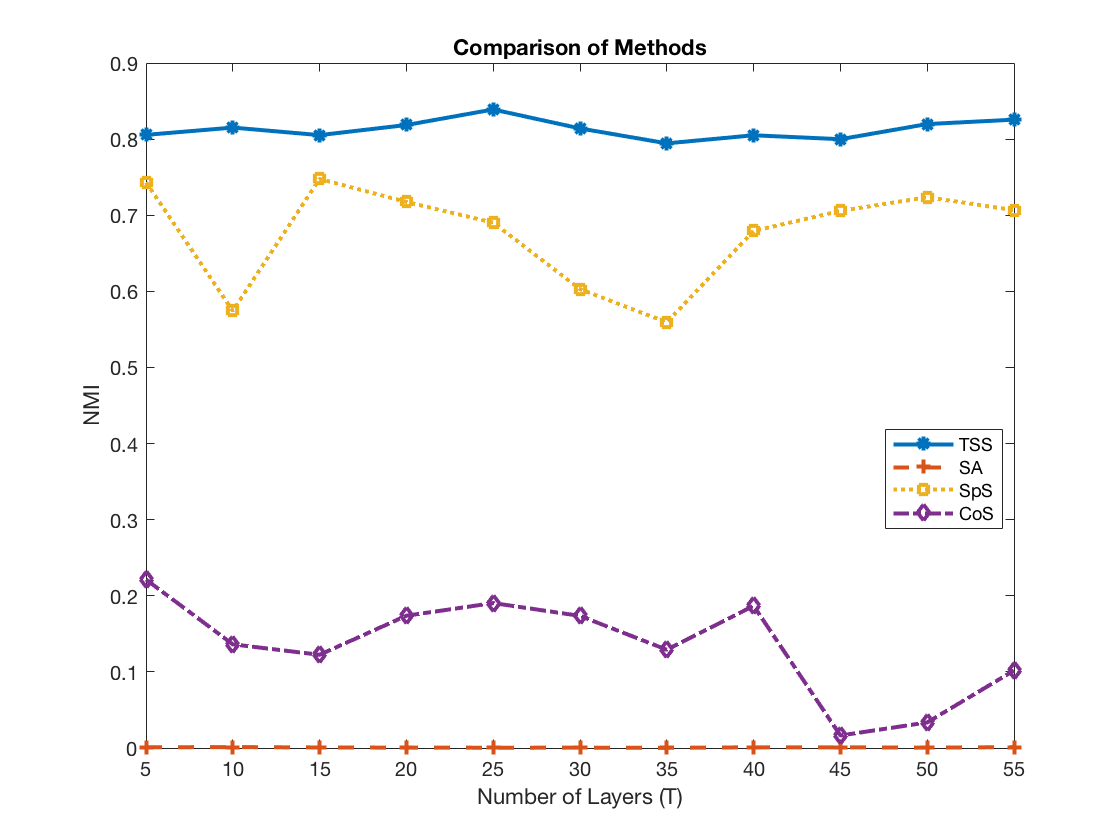}}
			\subfigure[]{\includegraphics[height=4.5cm,width=6cm]{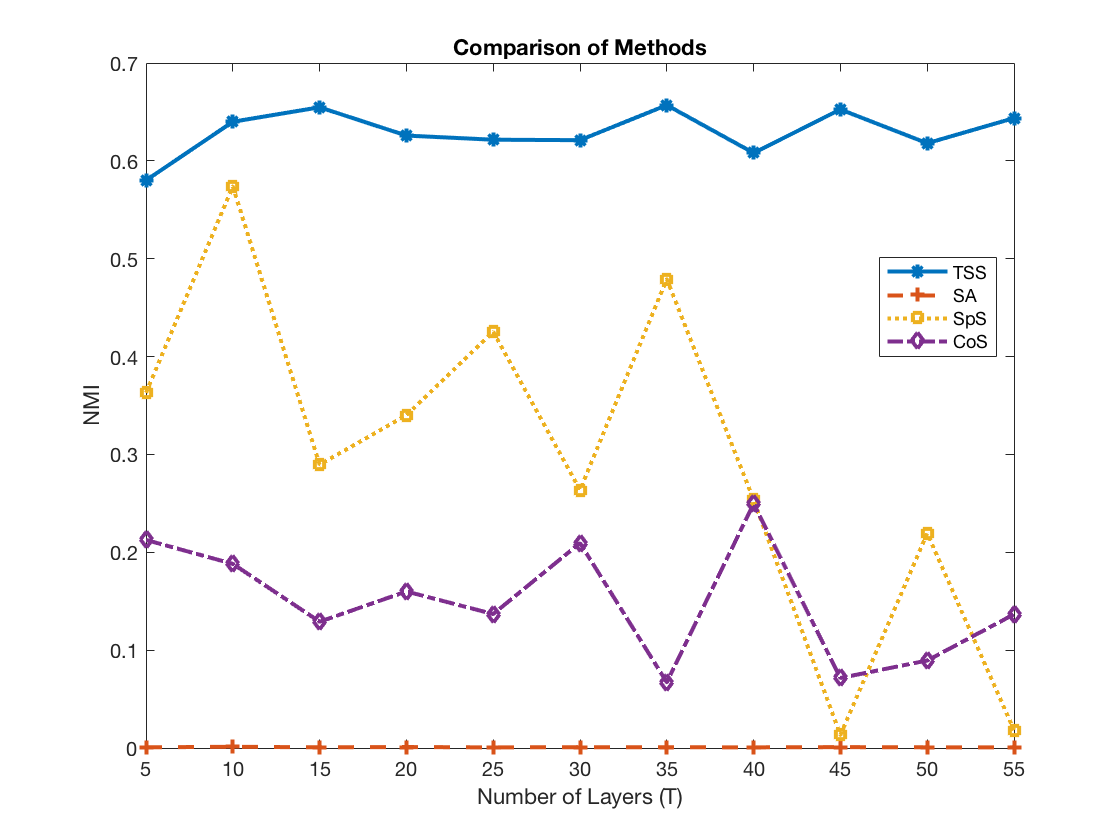}}
		\end{center}
		\caption{(a) NMI comparison using algorithms (i) as SA, (ii) as SpS, (iv) as CoS, and (v) as TSS; (b) NMI comparison using algorithms (ii) as SpS, (iii) as SA, (iv) as CoS, and (vi) as TSS. }
		\label{fig_msbm_mdcbm_disT}
	\end{figure}

	2. \emph{Changing number of layers ($T$) for multilayer SBM:} We vary number of layers $T$ from $5$ to $55$ with other parameters being $K=4$, $\vB^{(1)}_{4\times 4} = \frac{(\log n)^{4/3}}{n}\left(\V{J}_4 - \V{I}_4 \right)$, $\vB^{(2)}_{4\times 4} = \frac{(\log n)^{4/3}}{n}\V{J}_4$, $\vB^{(t)}_{4\times 4} = \frac{(\log n)^{4/3}}{nT}\V{J}_4$ for $t=3, \ldots, T$, $\V{\pi} = \frac{1}{4}\V{1}_{4\times 1}$, and $n=5000$. We compare the algorithms (i), (ii), (iv), and (v). The results on average NMI are given in Figure \ref{fig_msbm_mdcbm_disT}(a).
	\vspace{0.04in}
	
\begin{figure}[htp]
		\begin{center}
			\subfigure[]{\includegraphics[height=4.5cm,width=6cm]{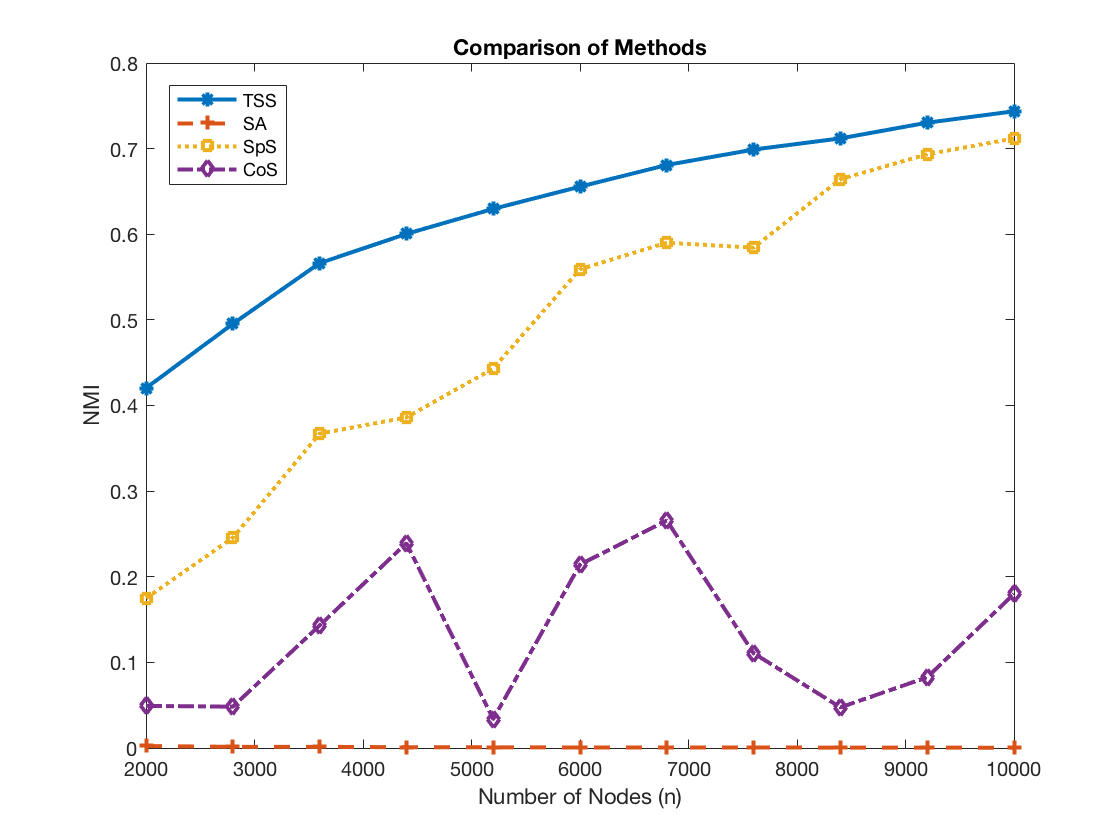}}
			%			\subfigure[]{\includegraphics[height=4cm,width=4cm]{plot_mdcbm_disass_KN1.png}}
		\end{center}
		\caption{(a) NMI comparison using algorithms (ii) - SpS, (iii) - SA, (iv) - CoS, and (vi) - TSS. }
		\label{fig_mdcbm_dis}
	\end{figure}

	3. \emph{Changing number of nodes ($n$) for multilayer DCBM:} We vary node size $n$ from $2000$ to $10000$ with other parameters being $K=4$, $\vB^{(1)}_{4\times 4} = \frac{(\log n)^{3/2}}{n}\left(\V{J}_4 - \V{I}_4 \right)$, $\vB^{(2)}_{4\times 4} = \frac{(\log n)^{3/2}}{n}\V{J}_4$, $\vB^{(t)}_{4\times 4} = \frac{(\log n)^{3/2}}{nT}\V{J}_4$ for $t=3, \ldots, T$, $\psi_i \stackrel{iid}{\sim} U(0.5, 1)$ for $i \in [n]$, and $T=11$. We compare the algorithms (ii), (iii), (iv), and (vi). The results on average NMI are given in Figure \ref{fig_mdcbm_dis}.
	\vspace{0.04in}
	
	4. \emph{Changing number of layers ($T$) for multilayer DCBM:} We vary number of layers $T$ from $5$ to $55$ with other parameters being $K=4$, $\vB^{(1)}_{4\times 4} = \frac{(\log n)^{3/2}}{n}\left(\V{J}_4 - \V{I}_4 \right)$, $\vB^{(2)}_{4\times 4} = \frac{(\log n)^{3/2}}{n}\V{J}_4$, $\vB^{(t)}_{4\times 4} = \frac{(\log n)^{3/2}}{nT}\V{J}_4$ for $t=3, \ldots, T$, $\V{\pi} = \frac{1}{4}\V{1}_{4\times 1}$, $\psi_i \stackrel{iid}{\sim} U(0.5, 1)$ for $i \in [n]$, and $n=5000$. We compare the algorithms (ii), (iii), (iv), and (vi). The results on average NMI are given in Figure \ref{fig_msbm_mdcbm_disT}(b).
	%\end{enumerate} 
	\vspace{0.1in}
	
	\noindent\emph{Scenario 3:} In this scenario, we consider a situation where $\cB=\{B^{(t)}: t\in [T]\}$ is a stochastic process. For each $t$ ($t\in [T]$), given, $B^{(t)}$, we simulate multilayer networks under the framework of \eqref{eq_sbm1} and \eqref{eq_dcbm1} of \textsection \ref{sec_model}. We consider two experiments under this scenario. Each experiment is repeated 25 times and the results are averages over the repetitions.
	\vspace{0.04in}
	
\begin{figure}[htp]
		\begin{center}
			\subfigure[]{\includegraphics[height=4.5cm,width=6cm]{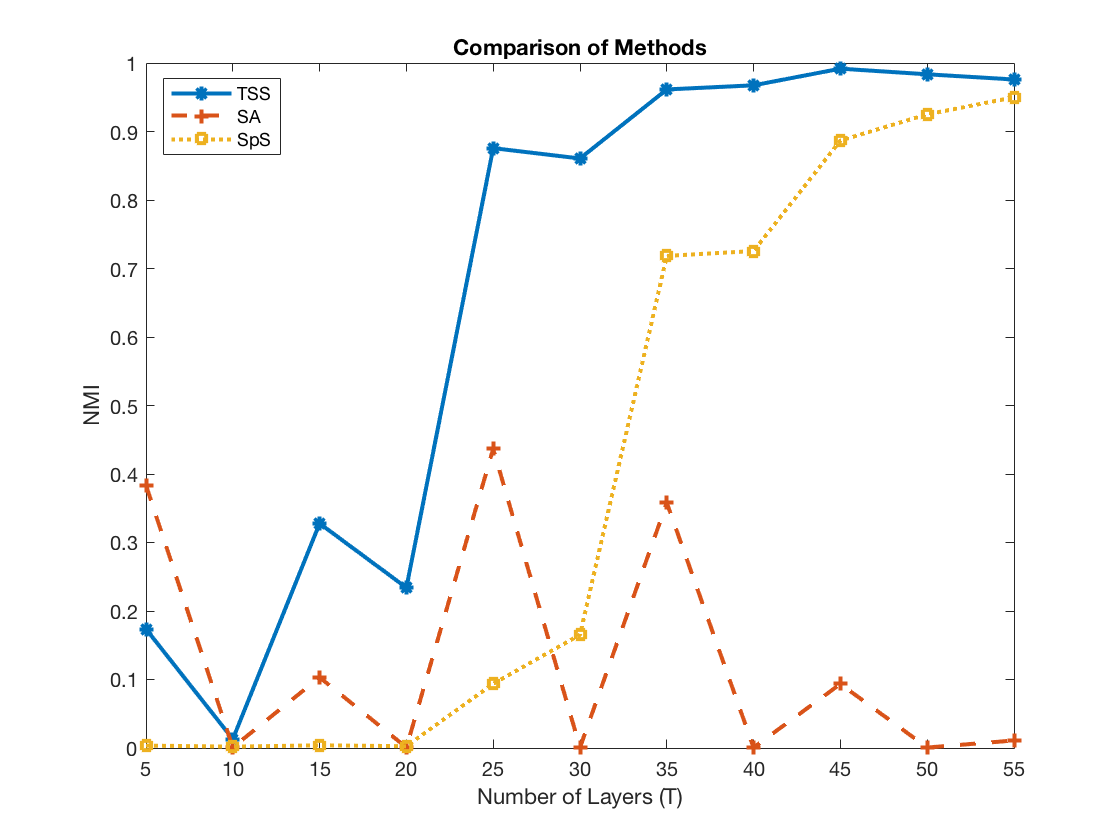}}
			\subfigure[]{\includegraphics[height=4.5cm,width=6cm]{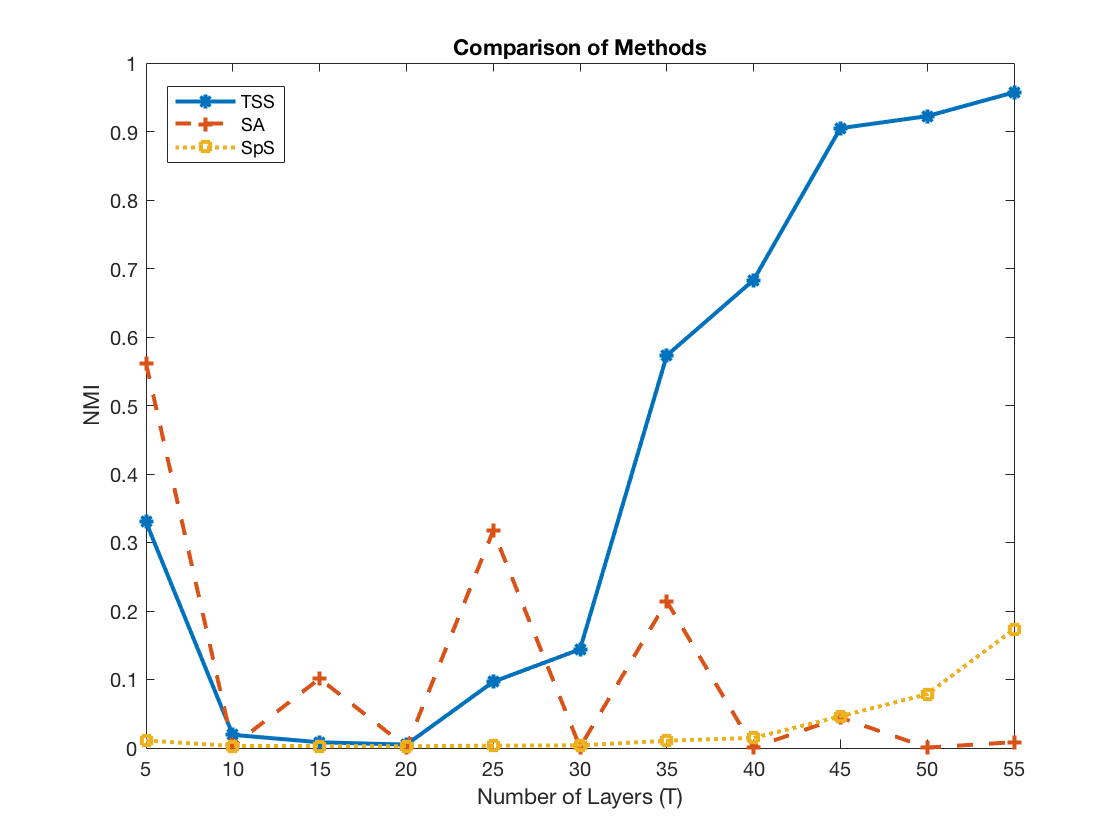}}
		\end{center}
		\caption{(a) NMI comparison using algorithms (i) as SA, (ii) as SpS, and (v) as TSS; (b) NMI comparison using algorithms (ii) as SpS, (iii) as SA, and (vi) as TSS. }
		\label{fig_msbm_mdcbm_depT}
	\end{figure}

	%\begin{enumerate}
	1. \emph{Changing number of layers ($T$) for multilayer SBM:} We vary number of layers $T$ from $5$ to $55$ with other parameters being $K=4$, $\V{\pi} = \frac{1}{4}\V{1}_{4\times 1}$, $n=5000$, $\vB^{(t)}_{4\times 4} = \frac{1}{n} \left(2\V{I}_4 + \ul b_t\V{J}_4 \right)$ for $t \in [5]$ where $\ul b_t = -7 + 7(t-1)/T$ for $t \in [T]$, and 
	\[\vB^{(t)}_{i,j} = \frac{20}{n(1+\exp(n\vB^{(t-5)}_{i, j} + \ul \eps_t))}\ \ \ \text{where, } \ul \eps_t \stackrel{iid}{\sim} N(0, 0.05).\]
	We compare the algorithms (i), (ii), and (v). The results on average NMI are given in Figure \ref{fig_msbm_mdcbm_depT}(a).
	\vspace{0.04in}
	
	2. \emph{Changing number of layers ($T$) for multilayer DCBM:} We vary number of layers $T$ from $5$ to $55$ with other parameters being same as previous experiment except $\ul b_t = -12 + 12(t-1)/T$ for $t \in [T]$. We compare the algorithms (ii), (iii), and (vi). The results on average NMI are given in Figure \ref{fig_msbm_mdcbm_depT}(b).
	%\end{enumerate} 
	\vspace{0.1in}
	
	%However under associative community structure spectral clustering using sum of adjacency matrices has a better recovery rate than sum of squared adjacency matrices.
	
	We see that Algorithm 1 works better in recovering community labels in all the scenarios compared to other algorithms for networks generated from multilayer stochastic block models as either $n \to \infty$ or $T \to \infty$. We also see that Algorithm 2 works better in recovering community labels in all the scenarios compared to other algorithms for networks generated from multilayer degree-corrected block models as either $n \to \infty$ or $T \to \infty$. Algorithm 1 and Algorithm 2 is also shown to recover community labels under dependent network layers. Algorithm 3 also recovers correct number of communities as $n \to \infty$. The simulation results are in concert with the theoretical results in Theorem \ref{ConsSum1}, Theorem \ref{ConsSum2}, Theorem \ref{Gen B^t}, and Theorem \ref{Khat_thm}.

	\section{Conclusion and Future Works}
	In this paper, we consider the problem of community detection for multi-relational networks with constant community memberships and changing connectivity matrices. We consider spectral clustering on aggregate versions of squared adjacency matrices. It is shown in the paper that under multilayer stochastic block model and multilayer degree-corrected block model, spectral clustering based on the sum of squared adjacency matrices has guarantee of consistent community recovery under weakest conditions on the degree parameters of the block models. We establish our claims both theoretically and empirically in the paper.
	% We also consider spherical spectral clustering based on the sum of adjacency matrices and squared adjacency matrices and give theoretical guarantee that the spherical spectral clustering method recovers community membership under dynamic degree-corrected block model too. 
	
	\subsection{Future Works}
	%Here we used bootstrap subsampling scheme to estimate \textit{local statistics} only. 
	Several extensions are possible from the current work. Some possible extensions of our work will include considering the cases where community memberships change with layers and the dependence of the network layers are more general, such as, dependence of probability of edge formation of a specific network layer on edge structure and community memberships of other network layers. Methods for community recovery with theoretical guarantee are quite rare for general multilayer networks and it would be good to investigate such problems in later works.
	
	\section{Acknowledgements}
	We thank Peter Bickel, Paul Bourgade, Ofer Zeitouni and Harrison Zhou for helpful discussions and comments.

	\bibliographystyle{imsart-nameyear}
	\bibliography{Biom1}
	
\end{document}